\documentclass[conference]{IEEEtran}

\IEEEoverridecommandlockouts

\usepackage{cite}
\usepackage{amsmath,amssymb,amsfonts}
\usepackage{algorithmic}
\usepackage{cprotect}
\usepackage{graphicx}
\usepackage{listings}
\usepackage{textcomp}
\usepackage{xcolor}
\def\BibTeX{{\rm B\kern-.05em{\sc i\kern-.025em b}\kern-.08em
    T\kern-.1667em\lower.7ex\hbox{E}\kern-.125emX}}

\usepackage[nolist]{acronym}
\usepackage[linesnumbered,ruled,vlined]{algorithm2e}
\usepackage{capt-of}
\usepackage{color, colortbl}
\usepackage{multirow}
\usepackage{pifont}
\usepackage{amsmath}
\usepackage[most]{tcolorbox}
\usepackage{tikz}
\usepackage[flushleft]{threeparttable}
\usepackage[hyphens]{url}

\graphicspath{{figures/}}

\SetAlCapNameFnt{\footnotesize}
\SetAlCapFnt{\footnotesize}

\definecolor{dkgreen}{rgb}{0,0.6,0}
\definecolor{gray}{rgb}{0.5,0.5,0.5}
\definecolor{mauve}{rgb}{0.58,0,0.82}
\newcommand{\llbox}[1]{
    \begin{tcolorbox}[
    sharp corners,
    boxsep = 0pt,
    left = 10pt,
    right = 10pt,
    colback=gray!10!white,
    colframe=blue!75!black,
    ]
    {#1}
    \end{tcolorbox}
}

\begin{acronym}
    \acro{cdg} [CDG] {Coverage Directed Test Generation}
\acro{ast} [AST] {Abstract Syntax Tree}
\acro{crv} [CRV] {Constrained Random Verification}
\acro{cve} [CVE] {Common Vulnerability Exposure}
\acro{dv} [DV] {Design Verification}
\acro{dut} [DUT] {Design Under Test}
\acro{e2e} [E2E] {End-to-End}
\acro{fsm} [FSM] {Finite State Machine}
\acro{gcp} [GCP] {Google Cloud Platform}
\acro{gcs} [GCS] {Google Cloud Storage}
\acro{hdl} [HDL] {Hardware Description Language}
\acro{hsb} [HSB] {Hardware Simulation Binary}
\acrodefplural{hsb} [HSBs] {Hardware Simulation Binaries}
\acro{hwfp} [HWFP] {Hardware Fuzzing Pipeline}
\acro{ip} [IP] {Intellectual Property}
\acro{loc} [LOC] {Lines of Code}
\acro{rtl} [RTL] {Register Transfer Level}
\acro{rot} [RoT] {Root-of-Trust}
\acro{soc} [SoC] {System-on-Chip}
\acro{sut} [SUT] {System Under Test}
\acro{sva} [SVA] {SystemVerilog Assertion}
\acro{tlul} [TL-UL] {TileLink Uncached Lightweight}
\acro{uvm} [UVM] {Universal Verification Methodology}
\end{acronym}

\newcommand{\needref}[1]{\textbf{\color{red}[Need Ref]}}

\newcommand{\hwf}{Hardware Fuzzing}
\newcommand{\etal}{{\textit{et al.}}\thinspace}
\newcommand{\numcircuits}{480}
\newcommand{\figline}{\vspace{3pt}\hrulefill}

\begin{document}

\title{Fuzzing Hardware Like Software
\thanks{
The project depicted is sponsored in part by the Defense Advanced Research Projects Agency, and the National Science Foundation under Grant CNS-1646130. The content of the information does not necessarily reflect the position or the policy of the Government, and no official endorsement should be inferred.
Approved for public release; distribution is unlimited.}
}

\author{
\IEEEauthorblockN{Timothy Trippel, Kang G. Shin}
\IEEEauthorblockA{\textit{Computer Science \& Engineering} \\
\textit{University of Michigan}\\
Ann Arbor, MI \\
\{trippel,kgshin\}@umich.edu}
\and
\IEEEauthorblockN{Alex Chernyakhovsky, Garret Kelly,\\ Dominic Rizzo}
\IEEEauthorblockA{\textit{OpenTitan} \\
\textit{Google, LLC}\\
Cambridge, MA \\
\{achernya,gdk,domrizzo\}@google.com}
\and
\IEEEauthorblockN{Matthew Hicks}
\IEEEauthorblockA{\textit{Computer Science} \\
\textit{Virginia Tech}\\
Blacksburg, VA \\
mdhicks2@vt.edu}
}

\maketitle
\thispagestyle{plain}
\pagestyle{plain}

\begin{abstract}
Hardware flaws are permanent and potent: hardware cannot be patched once fabricated, and any flaws may undermine even formally verified software executing on top. Consequently, verification time dominates implementation time. The gold standard in hardware Design Verification (DV) is concentrated at two extremes: random dynamic verification and formal verification. Both techniques struggle to root out the subtle flaws in complex hardware that often manifest as security vulnerabilities. The root problem with random verification is its undirected nature, making it inefficient, while formal verification is constrained by the state-space explosion problem, making it infeasible to apply to complex designs. What is needed is a solution that is directed, yet under-constrained.

Instead of making incremental improvements to existing hardware verification approaches, we leverage the observation that existing \textit{software fuzzers} already provide such a solution; we adapt it for hardware verification, thus leveraging existing---more advanced---software verification tools. Specifically, we translate RTL hardware to a software model and fuzz that model. The central challenge we address is how best to mitigate the differences between the hardware execution model and software execution model. This includes: 1) how to represent test cases, 2) what is the hardware equivalent of a crash, 3) what is an appropriate coverage metric, and 4) how to create a general-purpose fuzzing harness for hardware.

To evaluate our approach, we design, implement, and open-source a \textit{Hardware Fuzzing Pipeline} that enables fuzzing hardware at scale, using only open-source tools. Using our pipeline, we fuzz four IP blocks from Google's OpenTitan Root-of-Trust chip. Our experiments reveal a two orders-of-magnitude reduction in run time to achieve Finite State Machine (FSM) coverage over traditional dynamic verification schemes. Moreover, with our design-agnostic harness, we achieve over 88\% HDL line coverage in three out of four of our designs---even without any initial seeds.


\end{abstract}

\begin{IEEEkeywords}
Hardware Security, Design Verification, Fuzzing
\end{IEEEkeywords}


\section{Introduction}\label{section:introduction}
As Moore's Law~\cite{moore1965cramming} and Dennard scaling~\cite{dennard1974design} come to a crawl, hardware engineers must tailor their designs for specific applications 
in search of performance gains~\cite{chen2014diannao,
nowatzki2016pushing,jouppi2018motivation,hameed2010understanding,magaki2016asic}. 
As a result, hardware designs become increasingly unique and complex. 
For example, the Apple A11 Bionic \ac{soc}, released over three years ago in the iPhone 8, 
contains over 40 specialized \ac{ip} blocks, a number that doubles every four 
years~\cite{apple_soc_ip_cores}. Unfortunately, due to the state-explosion problem, 
\textbf{increasing design complexity increases \acf{dv} complexity, and therefore, 
the probability for design flaws to percolate into products.} 
Since 1999, 247 total \acp{cve} have been reported for Intel products, and of those, 
over 77\% (or 191) have been reported in the last four years~\cite{intel_cves}. 
While this may come as no surprise, given the onslaught of speculative execution 
attacks over the past few years~\cite{kocher2018spectre,lipp2018meltdown,
vanbulck2018foreshadow,ridl,canella2019fallout}, 
it highlights the correlation between hardware complexity and design flaws.

\begin{figure}[t]
\centering
\includegraphics[width=0.4\textwidth]{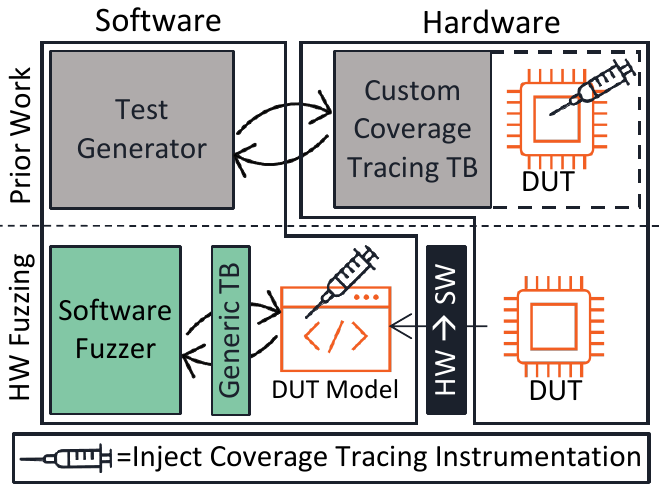}
\vspace*{-0.1in}
\caption{\footnotesize \textbf{Fuzzing Hardware Like Software.}
Unlike prior \acf{cdg} techniques~\cite{laeufer2018rfuzz,squillero2005microgp,gent2016fast,bose2001genetic}, we advocate for fuzzing software models of hardware directly, with a generic harness (testbench) and feature rich software fuzzers. In doing so, we address the barriers to realizing widespread adoption of \ac{cdg} in hardware \ac{dv}: 1) efficient coverage tracing, and 2) design-agnostic testing.
}
\label{fig:hwfuzz_overview}
\figline{}
\vspace*{-0.15in}
\end{figure}

Even worse, hardware flaws are \textit{permanent} and \textit{potent}. 
Unlike software, there is no general-purpose patching mechanism for hardware. 
Repairing hardware is both costly, and reputationally damaging~\cite{intel_bugs_cost}. 
Moreover, hardware flaws subvert even formally verified software that sits above~\cite{a2}. 
Therefore, detecting flaws in hardware designs \textit{before} fabrication and 
deployment is vital. Given these incentives, it is no surprise that hardware engineers 
often spend more time verifying their designs, than implementing 
them~\cite{wang2009electronic,fine2003coverage}.\footnote{It is estimated that up to 
70\% of hardware development time is spent verifying design 
correctness~\cite{fine2003coverage}.} 
Unfortunately, the multitude of recently-reported hardware vulnerabilities~\cite{kocher2018spectre,
lipp2018meltdown,vanbulck2018foreshadow,ridl,canella2019fallout,moghimi2020TPMFail} 
suggests current efforts are insufficient.

To address the threat of design flaws in hardware, engineers deploy two main \ac{dv} 
strategies: 1) \textit{dynamic} and 2) \textit{formal}. At one extreme, \textit{dynamic} 
verification involves driving concrete input sequences into a \ac{dut} during simulation, 
and comparing the \ac{dut}'s behavior to a set of invariants, or gold model. 
The most popular dynamic verification technique in practice today is known as 
\acf{crv}~\cite{cieplucha2019metric,ioannides2011introducing,yuan2003framework,uvm}. 
\ac{crv} attempts to decrease the manual effort required to develop simulation test
cases by randomizing input sequences in the hopes of \textit{automatically} maximizing 
exploration of the \ac{dut} state-space. At the opposite extreme, \textit{formal} 
verification involves proving/disproving properties of a \ac{dut} using mathematical 
reasoning like (bounded) model checking and/or deductive reasoning. 
While (random) \textit{dynamic} verification is effective at identifying 
surface flaws in even complex designs, it struggles to penetrate deep into a designs state space. 
In contrast, formal verification is effective at mitigating even 
deep flaws in small hardware designs, but fails, in practice, against larger designs.

In search of a hybrid approach to bridge these \ac{dv} extremes, researchers have 
ported software testing techniques to the hardware domain in hopes of improving hardware test 
generation to maximize coverage. In the hardware domain, these approaches 
are referred to as \ac{cdg}~\cite{fine2003coverage,cieplucha2019metric,ioannides2011introducing,teplitsky2015coverage,guzey2007coverage,wang2018accelerating,laeufer2018rfuzz,zhang2018recursive,zhang2018end,bose2001genetic}. 
Like their software counterparts, \ac{cdg} techniques deploy coverage metrics---e.g., \ac{hdl}
line, \ac{fsm}, functional, etc.---in a feedback loop to generate
tests that further increase state exploration. 

While promising, why has \ac{cdg} not seen widespread adoption in hardware \ac{dv}? 
As Laeufer \etal{} point out~\cite{laeufer2018rfuzz}, this is likely fueled by several \textbf{key technical challenges, resulting from dissimilarities between software and hardware execution models}. First, unlike software, \ac{rtl} hardware is not inherently executable. Hardware designs must be simulated, after being translated to a software model and combined with a design-specific testbench and simulation engine, to form a \acf{hsb} (Fig.~\ref{fig:hw_sim_binary}).
This level of indirection, increases both the complexity and computational effort in tracing test coverage of the hardware.
Second, unlike most software, hardware requires \textit{sequences} of structured inputs to drive meaningful state transitions, that must be tailored to each \ac{dut}. For example, while most software often accepts input in the form of a fixed set of file(s) that contain a loosely-structured set of bytes (e.g., a JPEG or PDF), hardware often accepts input from an ongoing stream of bus transactions. Together, these challenges have resulted in \ac{cdg} approaches that implement custom: 1) coverage-tracing techniques that still suffer from poor scalability~\cite{laeufer2018rfuzz,ioannides2011introducing}, and 2) test generators that have limited compatibility to a small class of \acp{dut}, e.g., processors~\cite{squillero2005microgp,zhang2018end,bose2001genetic}.

To supplement traditional dynamic verification methods, we propose an alternative \ac{cdg} technique we call \hwf{}. \textbf{Rather than translating software testing methods to the hardware domain, we advocate for translating hardware designs to software models} and fuzzing those models directly (Fig.~\ref{fig:hwfuzz_overview}). While fuzzing hardware in the software domain eliminates coverage-tracing bottlenecks of prior \ac{cdg} techniques~\cite{laeufer2018rfuzz,squillero2005microgp,ioannides2011introducing}, since software can be instrumented at compile time to trace coverage, it does not inherently solve the design compatibility issue. Moreover, it creates other challenges we must address. Specifically, to fuzz hardware like software, we must adapt software fuzzers to:
\begin{enumerate}
    \item interface with \acp{hsb} that: a) contain other components besides the \ac{dut}, and b) 
    require unique initialization.
    \item account for differences between how hardware and software process inputs, and its impact on 
    exploration depth.
    \item design a general-purpose fuzzing harness and a suitable grammar that ensures meaningful mutation.
\end{enumerate}

To address these challenges, we first propose (and evaluate) strategies for interfacing software 
fuzzers with \acp{hsb} that optimize performance and trigger the \ac{hsb} to crash upon detection of 
incorrect hardware behavior. 
Second, we show that maximizing code coverage of the \ac{dut}'s software model, by construction, maximizes hardware code coverage.
Third, we design an interface to map fuzzer-generated test-cases 
to hardware input ports. Our interface is built on the observation that unlike most 
software, hardware requires piecing together a sequence of inputs to effect meaningful 
state transitions. 
Lastly, we propose a new interface for fuzzing hardware in a design agnostic manner: 
the \textit{bus interface}. Moreover, we design and implement a generic harness, and 
create a corresponding grammar that ensures meaningful mutations to fuzz bus transactions.
Fuzzing at the bus interface solves the final hurdle to realizing 
widespread deployability of \ac{cdg} in hardware \ac{dv}, as it enables us to reuse the same 
testbench harness to fuzz any \ac{rtl} hardware that speaks the same bus protocol, irrespective of 
the \ac{dut}'s design or implementation.

To demonstrate the effectiveness of our approach, we design, implement, and open-source 
a \acf{hwfp}~\cite{hwfp}, inspired by Google's OSS-Fuzz~\cite{oss_fuzz}, capable of fuzzing \ac{rtl} 
hardware at scale (Fig.~\ref{fig:hwfp}). Using our \ac{hwfp} we compare \hwf{} against a 
conventional \ac{crv} technique when verifying over \numcircuits{} variations of a 
sequential \ac{fsm} circuit. Across our experiments, we observe over two orders-of-magnitude 
reduction in time to reach full \ac{fsm} coverage by fuzzing hardware like software. 
Moreover, using our bus-specific hardware fuzzing grammar, we fuzz four commercial \ac{ip} 
cores from Google's OpenTitan silicon \ac{rot}~\cite{opentitan}. Even without seeding the 
fuzzer, we achieve over 88\% \ac{hdl} line coverage after only 1-hour of fuzzing 
on three of the four cores.

In summary, we:
\begin{itemize}
    \item propose deploying feature-rich software fuzzers as a \ac{cdg} approach to 
    solve inefficiencies in hardware \ac{dv} (\S\ref{section:fuzzing_hw_like_sw});
 
    \item provide empirically-backed guidance on how to: 1) isolate the \ac{dut} portion 
    of \acp{hsb}, and 2) minimize overhead of persistent hardware resets, for 
    fuzzing (\S\ref{subsubsection:interfacing_swfuzzers_w_hw_design} \& \S\ref{subsection:fuzzing_opts_eval});
       
    \item develop a technique to map fuzzer-generated testcases across both space and time 
    to create a \textit{sequence} of inputs to stimulate software models of hardware (\S\ref{subsubsection:processing_fuzzer_generated_tests});
    
    \item design and evaluate several bus-specific \hwf{} harnesses and grammars to 
    facilitate fuzzing all bus-based hardware cores (\S\ref{subsubsection:hwf_bus_harness}, 
    \S\ref{subsubsection:hwf_bus_grammar} \& \S\ref{subsection:hwf_grammar_eval});

    \item design, implement, and open-source a \ac{hwfp}~\cite{hwfp} that 
    continuously fuzzes \ac{rtl} hardware at scale on \ac{gcp} 
    (\S\ref{section:implementation}); and
    
    \item demonstrate \hwf{} provides two orders-of-magnitude reduction in run time to 
    achieve comparable (or better) \ac{fsm} coverage to (or than) current state-of-the-art 
    \ac{crv} schemes (\S\ref{subsection:fuzzing_vs_crv_eval}).
\end{itemize}

\section{Background}\label{section:background}
There are two main hardware verification methods: 1) \textit{dynamic} and 
2) \textit{formal}. While there have been significant advancements in deploying 
formal methods in \ac{dv} workflows~\cite{kaivola2009replacing,opentitan,zhang2018end}, 
dynamic verification remains the gold standard due to its scalability towards complex 
designs~\cite{laeufer2018rfuzz}. Therefore, we focus on improving \textit{dynamic} 
verification by leveraging advancements in the software fuzzing community. Below, we 
provide a brief overview of the current state-of-the-art in dynamic hardware 
verification, and software fuzzing.

\subsection{Dynamic Verification of Hardware}
\label{subsection:background_dynamic_verification}
\textit{Dynamic} verification of hardware typically involves three steps: 
1) \textbf{test generation}, 2) \textbf{hardware simulation}, and 3) \textbf{test evaluation}.
First, during \textit{test generation}, a sequence of inputs are crafted to stimulate 
the \ac{dut}. Next, the \ac{dut}'s behavior---in response to the input sequence---is 
simulated during \textit{hardware simulation}. Lastly, during \textit{test 
evaluation}, the \ac{dut}'s simulation behavior is checked for correctness. 
These three steps are repeated until all interesting \ac{dut} behaviors have been 
explored. How do we know when we have explored all interesting behaviors? 
To answer this question, verification engineers measure coverage of both: 1) manually 
defined functional behaviors (functional coverage) and 2) the \ac{hdl} implementation 
of the design (code coverage) \cite{tasiran2001coverage,jou1999coverage,piziali2007functional}.

\subsubsection{\textbf{Test Generation}}
To maximize efficiency, \ac{dv} engineers aim to generate as few test vectors as possible 
that still close coverage. To achieve this goal, they deploy two main test generation 
strategies: 1) constrained-random and 2) coverage-directed. The former is typically 
referred to holistically as \textit{\acf{crv}}, and the latter as \textit{\acf{cdg}}. 
\ac{crv} is a partially automated test generation technique where manually-defined input 
sets are randomly combined into transaction sequences~\cite{yuan2003framework,uvm}. 
While better than an \textit{entirely} manual approach, \ac{crv} still requires some 
degree of manual tuning to avoid inefficiencies, since the test generator has no 
knowledge of test coverage. Regardless, \ac{crv} remains a popular dynamic verification 
technique today, and its principles are implemented in two widely deployed (both 
commercially and academically) hardware \ac{dv} frameworks: 1) Accellera's \ac{uvm} 
framework (SystemVerilog)~\cite{uvm} and 2) the open-source cocotb (Python) 
framework~\cite{cocotb}. 

To overcome \ac{crv} shortcomings, researchers have proposed \ac{cdg}~\cite{fine2003coverage,cieplucha2019metric,ioannides2011introducing,teplitsky2015coverage,guzey2007coverage,wang2018accelerating,laeufer2018rfuzz,zhang2018recursive,zhang2018end,bose2001genetic,gent2016fast,squillero2005microgp}, or using test coverage 
feedback to drive future test generation. Unlike \ac{crv}, \ac{cdg} does not randomly 
piece input sequences together in hopes of exploring new design state. Rather, it 
\textit{mutates} prior input sequences that explore uncovered regions of the design to 
iteratively expand the coverage boundary. Unfortunately, due to 
deployability challenges, e.g., slow coverage tracing and limited applicability to a small
set of \acp{dut}, \ac{cdg} has not seen widespread adoption in 
practice~\cite{laeufer2018rfuzz}. In this paper, we recognize that existing software 
fuzzers provide a solution to many of these deployability challenges,
and therefore advocate for verifying hardware using software verification tools. 
The central challenges in making this possible are adapting software fuzzers to verify hardware,
widening the scope of supported designs, and increasing automation of verification.

\begin{figure}[t]
\centering
\includegraphics[width=0.3\textwidth]{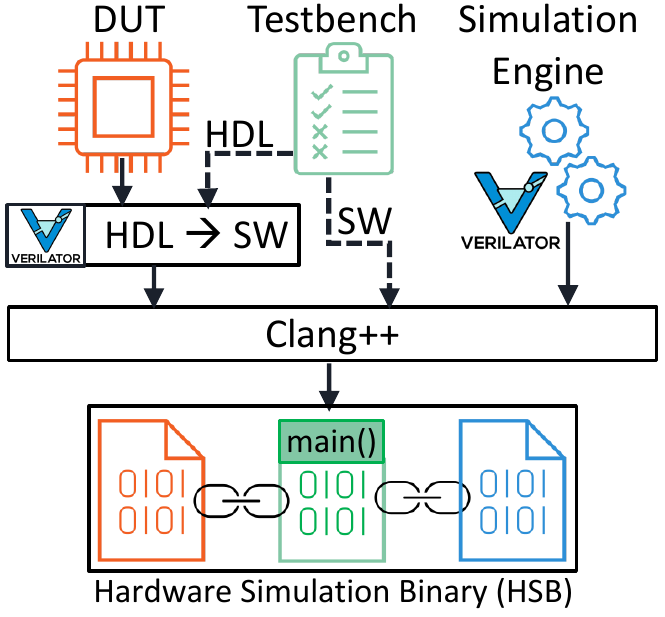}
\vspace*{-0.1in}
\cprotect\caption{\footnotesize \textbf{\acf{hsb}.}
To simulate hardware, the \ac{dut}'s \ac{hdl} is first translated to a software model, and then compiled/linked with a testbench (written in \ac{hdl} or software) and simulation engine to form a \textit{\acf{hsb}}. Executing this binary with a sequence of test inputs simulates the behavior of the \ac{dut}.
}
\label{fig:hw_sim_binary}
\figline{}
\vspace*{-0.15in}
\end{figure}
\begin{figure*}[t]
\centering
\includegraphics[width=0.75\textwidth]{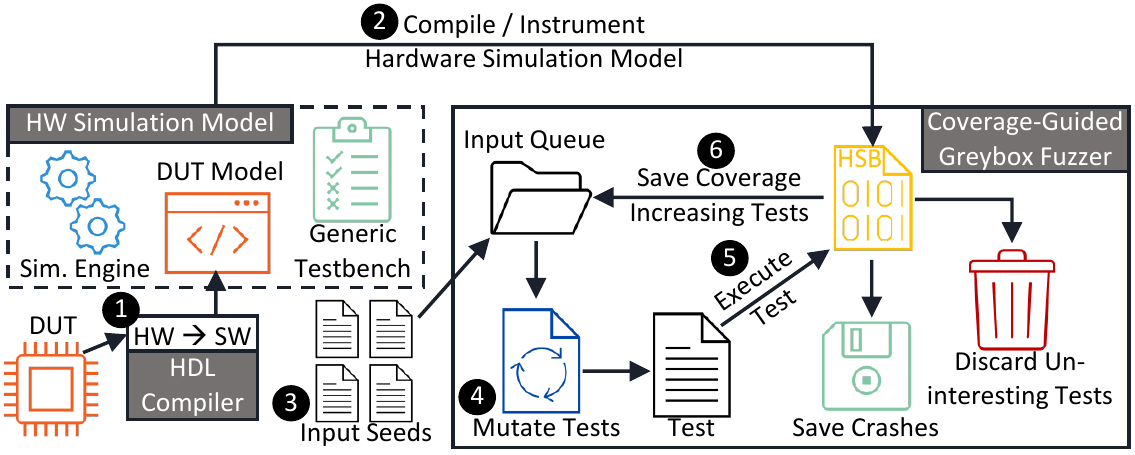}
\vspace*{-0.05in}
\caption{\footnotesize \textbf{\hwf{}.}
Fuzzing hardware in the software domain involves: translating the hardware \ac{dut} to a functionally equivalent software model (1) using a SystemVerilog compiler~\cite{verilator}, compiling and instrumenting a \acf{hsb} to trace coverage (2), crafting a set of seed input files (3) using our design-agnostic grammar (\S~\ref{subsubsection:hwf_bus_grammar}), and fuzzing the \ac{hsb} with a coverage-guided greybox software fuzzer~\cite{afl,libfuzzer,honggfuzz} (4--6).
}
\label{fig:hwf_detailed_process}
\figline{}
\vspace*{-0.15in}
\end{figure*}

\subsubsection{\textbf{Hardware Simulation}}\label{subsubsection:background_hw_simulation}
While there are several commercial \cite{vcs,modelsim,xcelium} and open-source 
\cite{verilator,icarus} hardware simulators, most work in the same general manner, 
as shown in Fig.~\ref{fig:hw_sim_binary}. First, they translate hardware 
implementations (described in \ac{hdl}) into a software model, usually in C/C++. 
Next, they compile the software 
model and a testbench---either translated from \ac{hdl}, or implemented in software 
(C/C++)---and link them with a simulation engine. Together, all three components form an 
\textit{\acf{hsb}} (Fig.~\ref{fig:hw_sim_binary}) that can be executed to simulate the 
design. Lastly, the \ac{hsb} is executed with the inputs from the testbench to capture the design's 
behavior. Ironically, even though commercial simulators convert the hardware to software,
they still \textbf{rely on hardware-specific verification tools, likely because 
software-oriented tools fail to work on hardware models---without the lessons in this paper}.
To fuzz hardware in the software domain, we take advantage of the transparency 
in how an open-source hardware simulator, Verilator~\cite{verilator}, generates an 
\ac{hsb}. Namely, we \textit{intercept} the software model of the hardware after 
translation, and instrument/compile it for coverage-guided fuzzing 
(Fig.~\ref{fig:hwf_detailed_process}).

\subsubsection{\textbf{Test Evaluation}}\label{subsubsection:background_test_evaluation}
After simulating a sequence of test inputs, the state of the hardware (both internally 
and its outputs) are evaluated for correctness. There are two main approaches for 
verifying design correctness: 1) invariant checking and 2) (gold) model 
checking. In invariant checking, a set of assertions (e.g., \acp{sva} or software side 
C/C++ assertions) are used to check properties of the design have not been violated. 
In model checking, a separate model of the \ac{dut}'s correct behavior is emulated in
software, and compared to the \ac{dut}'s simulated behavior.
We support such features and adopt both invariant violations and 
golden model mismatches as an analog for software crashes in our hardware fuzzer.

\subsection{Software Fuzzing}
Software fuzzing is an automated testing technique designed to identify security 
vulnerabilities in software~\cite{sutton2007fuzzing}. Thanks to its success, 
it has seen widespread adoption in both industry~\cite{bounimova2013billions} 
and open-source~\cite{oss_fuzz} projects. In principle, 
fuzzing typically involves the following three main steps~\cite{nagy2019full}:
1) \textbf{test generation}, 2) \textbf{monitoring test execution}, and
3) crash triaging. During test generation, program inputs are synthesized to 
exercise the target binary. 
Next, these inputs are fed to the program under test, and its execution is monitored. 
Lastly, if a specific test causes a crash, that test is further analyzed to find the 
root cause. This process is repeated until all, or most, of the target binary has 
been explored. Below we categorize fuzzers by how they implement the first two steps.

\subsubsection{\textbf{Test Generation}}
Most fuzzers generate test cases in one of two ways, using: 1) a grammar, or 2) mutations.
Grammar-based fuzzers~\cite{dharma,gramfuzz,aschermann2019nautilus,peachfuzz,wang2017skyfire,wang2019superion} 
use a human-crafted grammar to constrain tests to comply with structural requirements of 
a specific target application. Alternatively, mutational fuzzers take a correctly 
formatted test as a seed, and apply mutations to the seed to create new tests. 
Moreover, mutational fuzzers are tuned to be either: 1) \textit{directed}, or 
2) \textit{coverage-guided}. Directed mutational fuzzers
\cite{bohme2017coverage,zongfuzzguard,chen2018hawkeye,aschermannijon,osterlundparmesan,you2017semfuzz,wang2010taintscope} 
favor mutations that explore specific region within the target binary, 
i.e., prioritizing exploration \textit{location}. Conversely, coverage-guided mutational 
fuzzers~\cite{afl,libfuzzer,honggfuzz,serebryany2016continuous,syzkaller,rawat2017vuzzer} 
favor mutations that explore as much of the target binary as possible, 
i.e., prioritizing exploration \textit{completeness}. 
For this work, we favor the use of mutational, coverage-guided fuzzers, as 
they are both design-agnostic, and regionally generic.

\subsubsection{\textbf{Test Execution Monitoring}}
Fuzzers monitor test execution using one of three approaches: 1) blackbox, 
2) whitebox, or 3) greybox. 
Fuzzers that only monitor program inputs and outputs are classified as \textit{blackbox} 
fuzzers~\cite{autodafe,dharma,peachfuzz}. Alternatively, fuzzers that track
detailed execution paths through programs with fine-grain program analysis (source code required) and 
constraint solving are known as \textit{whitebox} 
fuzzers~\cite{stephens2016driller,cadar2008klee,cha2012unleashing,huangpangolin,yun2018qsym,chipounov2011s2e,godefroid2012sage,wang2010taintscope}. 
Lastly, \textit{greybox} 
fuzzers~\cite{afl,bohme2017coverage,zongfuzzguard,honggfuzz,serebryany2016continuous,aschermann2019nautilus,osterlundparmesan,you2017semfuzz,wang2017skyfire,wang2019superion,syzkaller,peng2018t,triforce_afl,rawat2017vuzzer}
offer a trade-off between black- and whitebox fuzzers by deploying lightweight program 
analysis techniques, such as code-coverage tracing. 
Since Verilator~\cite{verilator} produces raw C++ source code from \ac{rtl} hardware, our approach can leverage \textit{any} software fuzzing technique---white, grey, or blackbox. In our current implementation, we deploy greybox fuzzing, due to its popularity in the software testing community.

\section{\hwf{}}\label{section:fuzzing_hw_like_sw}
To take advantage of advances in software fuzzing for hardware \ac{dv}, we propose translating
hardware designs to software models, and fuzzing the model directly. We call this approach,
\textbf{\hwf{}}, and illustrate the process in Fig.~\ref{fig:hwf_detailed_process}.
Below, we first motivate our approach by describing how hardware is already translated to 
the software domain for simulation, and 
that software fuzzers provide a solution to a key technical challenge in \ac{cdg}:
scalable coverage tracing. 
Then, we pose several challenges in adapting software fuzzers to fuzz \acp{hsb} 
(in a design-agnostic fashion), and present solutions to overcome these challenges.

\subsection{Why Fuzz Hardware like Software?}
We observe two key benefits of fuzzing hardware in the software domain. First, 
hardware is already translated to a software model for simulation purposes 
(\S\ref{subsubsection:background_hw_simulation}). Second, unlike prior \ac{cdg} 
approaches~\cite{laeufer2018rfuzz,squillero2005microgp}, we recognize that software 
fuzzers already provide an efficient solution for tracing coverage. 
Below we explain how \ac{rtl} hardware is translated to executable software, and 
why software fuzzers implicitly maximize hardware coverage by generating tests 
that maximize coverage of the \ac{hsb}.

\subsubsection{\textbf{Translating \acs{hdl} to Software}}\label{subsection:translating_hw_to_sw}
Today, simulating \ac{rtl} hardware involves translating \ac{hdl} into a functionally 
equivalent software (C/C++) model that can be compiled and executed 
(\S\ref{subsubsection:background_hw_simulation}). To accomplish this, most hardware 
simulators~\cite{icarus,verilator} contain an \ac{rtl} compiler to perform the translation. 
Therefore, we leverage a popular open-source hardware simulator, Verilator~\cite{verilator}, 
to translate SystemVerilog \ac{hdl} into a cycle-accurate C++ model for fuzzing.

Like many compilers, Verilator first performs lexical analysis and parsing (of the \ac{hdl}) 
with the help of Flex~\cite{paxson2007lexical} and Bison~\cite{bison}, to generate an \ac{ast}. 
Then, it performs 
a series of passes over the \ac{ast} to resolve parameters, propagate constants, 
replace \textit{don't cares} (Xs) with random values, eliminate dead code, unroll 
loops/generate statements, and perform several other optimizations. 
Finally, Verilator generates C++ (or SystemC) code representing a cycle-accurate model 
of the hardware. It creates a C++ class for each Verilog module, and organizes 
classes according to the original \ac{hdl} module hierarchy~\cite{zhang2018end}. 

To interface with the model, Verilator exposes public member variables for each input/output 
to the top-level module, and a public \verb|eval()| method (to be called in a loop) in 
the top C++ class. Each input/output member variable is mapped to single/arrayed \verb|bool|, 
\verb|uint32_t|, or \verb|uint64_t| data types, depending on the width of each signal.
Each call to \verb|eval()| updates the model based on the current values assigned to 
top-level inputs and internal states variables. Two calls represent a single clock cycle 
(one call for each rising and falling clock edges).

\subsubsection{\textbf{Tracing Hardware Coverage in Software}}
To efficiently explore a \ac{dut}'s state space, \ac{cdg} techniques rely on tracing 
coverage of past test cases to generate future test cases. There are two main categories 
of coverage metrics used in hardware verification~\cite{tasiran2001coverage,jou1999coverage,piziali2007functional}: 
1) \textit{code coverage}, and 2) \textit{functional coverage}. 
The coarsest, and most used, code coverage metric is \textit{line coverage}. 
Line coverage measures the percentage of \ac{hdl} lines that have been exercised during simulation. 
Alternatively, \textit{functional coverage} measures the percentage of various high-level design 
functionalities---defined using special \ac{hdl} constructs like SystemVerilog Coverage 
Points/Groups---that are exercised during simulation. Regardless of the coverage metric used, 
tracing \ac{hdl} coverage during simulation is often slow, since coverage traced in the 
software (simulation) domain must be mapped back to the hardware domain~\cite{jou1999coverage}.

In an effort to compute \ac{dut} coverage efficiently, and in an \ac{hdl}-agnostic manner, 
prior \ac{cdg} techniques develop custom coverage metrics, e.g., \textit{multiplexer coverage}~\cite{laeufer2018rfuzz}, 
that can be monitored by instrumenting the \ac{rtl} directly. 
However, this approach has two drawbacks. First, the hardware must be simulated 
on an FPGA (simulating within software is just as slow). Second, the authors provide 
no indication that their custom coverage metrics actually translate to coverage metrics 
\ac{dv} engineers care about.

Rather than make incremental improvements to existing \ac{cdg} techniques, we recognize that: 
1) software fuzzers provide an efficient mechanism---e.g., binary instrumentation---to trace 
coverage of compiled C++ hardware models (\acp{hsb}), and 2) characteristics of how Verilator 
translates \ac{rtl} hardware to software makes mapping software coverage to hardware 
coverage implicit. On the software side, there are three main code coverage metrics of increasing granularity: 1) basic block, 2) basic block edges, and 3) basic block paths~\cite{nagy2019full}. The most popular coverage-guided fuzzers---AFL~\cite{afl}, libFuzzer~\cite{libfuzzer}, and honggfuzz~\cite{honggfuzz}---all trace \textit{edge} coverage. On the hardware side, Verilator conveniently generates straight-line C++ code for both blocking and non-blocking\footnote{Verilator imposes an order on the non-blocking assignments since C++ does not have a semantically equivalent assignment operator~\cite{verilator,zhang2018end}. Regardless, this ordering does not effect code coverage.} SystemVerilog statements~\cite{zhang2018end}, and injects conditional code blocks (basic blocks) for SystemVerilog Assertions and Coverage Points. Therefore, optimizing test-generation \textit{edge} coverage of the software model of the hardware during simulation, translates to optimizing both \textit{code} and \textit{functional} coverage of the hardware itself. We demonstrate this artifact in \S\ref{subsubsection:hwf_grammar_eval_results} of our evaluation.

\subsection{Adapting Software Fuzzers to Fuzz Hardware}
While software fuzzers contain efficient mechanisms for tracing coverage of 
\acp{hsb}---e.g., binary instrumentation---interfacing them with \acp{hsb}, in a 
design-agnostic manner is non-trivial. 
Below, we highlight several challenges in fuzzing \acp{hsb} with software fuzzers, 
and propose solutions to overcome them.

\subsubsection{\textbf{Interfacing Software Fuzzers with \acp{hsb}}}
\label{subsubsection:interfacing_swfuzzers_w_hw_design}
Na\"{i}vely, a \ac{dv} engineer may interface the \ac{hsb} directly with a software fuzzer 
(like~\cite{afl,libfuzzer,honggfuzz}) by compiling the \ac{hsb} source code alongside the 
testbench harness (Algo.~\ref{algo:fuzzer_tb_harness}) and simulation engine with one of the 
fuzzer-provided wrappers for Clang. However, they would be ignoring two key differences between 
typical software applications and \acp{hsb} that may degrade fuzzer performance. 
First, \acp{hsb} have other components---a testbench and simulation engine 
(Fig.~\ref{fig:hw_sim_binary})---that are not part of the \ac{dut}. While the \ac{dut} is 
manipulated through the testbench and simulation engine, instrumenting all 
components \acp{hsb} actually degrades fuzzer performance 
(\S\ref{subsubsection:instrumenting_hw_for_fuzzing}). 
Additionally, unlike software, the \ac{dut} software model must be reset and initialized, 
prior to processing any inputs. Depending on the size of the \ac{dut}, 
this process can require special configuration of the testbench, 
i.e., initializing the fuzzer to snapshot the hardware simulation process \textit{after} 
reset and initialization of the \ac{dut} (\S\ref{subsubsection:hw_resets_opt}).

\subsubsection{\textbf{Interpreting Fuzzer-Generated Tests}}
\label{subsubsection:processing_fuzzer_generated_tests}
For most software, a single input often activates an entire set of state transitions within the program. 
Consequently, the most popular software fuzzers assume the target binary reads a single dimensional 
input---e.g., a single image or document---from either a file, \verb|stdin|, or a byte 
array~\cite{afl,honggfuzz,libfuzzer}. Unfortunately, the execution model of hardware is different. 
In an \ac{hsb}, a \textit{sequence} of inputs is required to activate 
state transitions within the \ac{dut}. For example, a 4-digit lock (with a keypad) only has a
\textit{chance} of unlocking if a sequence of four inputs (test cases) are provided. Fuzzing this
lock with single test cases (digits), will fail. Likewise, fuzzing \acp{hsb} with software fuzzers
that employ a \textit{single-test-case-per-file} model will also fail.
Therefore, to stimulate hardware with software fuzzers, we propose a new interface for interpreting 
single dimensional fuzzer-generated tests in two dimensions: space and time. We implement 
this interface in the form of a generic fuzzing harness (testbench)---shown in 
Algo.~\ref{algo:fuzzer_tb_harness}---that continuously: 1) reads byte-level portions of 
fuzzer-generated test files, 2) maps these bytes to hardware input ports, and 3) advances 
the simulation clock by calling the model's \verb|eval()| method twice, until there are 
no remaining bytes to process. With our fuzzing harness, we transform one-dimensional
test inputs, into a two-dimensional \textit{sequence} of inputs.

\begin{algorithm}[ht]
\footnotesize
\label{algo:fuzzer_tb_harness}
\caption{Generic \hwf{} harness (testbench) that maps one-dimensional fuzzer-generated test files to both spatial and temporal dimensions.}
\SetAlgoLined
\KwIn{fuzz\_test\_file.hwf}
$dut \leftarrow Vtop()$;\\
$tf \leftarrow open(fuzz\_test\_file.hwf)$;\\
\While {tf not empty}{
    \ForEach{$port \in dut.inputs$}{
        tf.read((uint\_8t*) port, $sizeof$(port));\\
        \For{$k \gets 1$ to $2$}{
            $clock \leftarrow (clock\,+\,1)\,\%\,2$;\\
            $dut.eval()$;\\
        }
    }
}
\end{algorithm}

\subsubsection{\textbf{Bus-Centric Harness}}
\label{subsubsection:hwf_bus_harness}
While the multi-dimensional fuzzing interface we develop enables fuzzer-generated 
tests to effect state transitions in hardware, it is not design-agnostic. Specifically,
the ports of a hardware model are not iterable (Algo.~\ref{algo:fuzzer_tb_harness}: line 4).
A \ac{dv} engineer would have to create a unique fuzz harness (testbench) for each \ac{dut} they verify.
To facilitate \ac{dut} portability, we take inspiration from how hardware engineers 
interface \ac{ip} cores within an \ac{soc}~\cite{ot_comportability_guide}.
Specifically, we propose fuzzing \ac{ip} cores at the bus interface using a
bus-centric harness. 

To implement this harness, we could alter our prior harness (Algo.~\ref{algo:fuzzer_tb_harness})
by mapping bytes from fuzzer-generated test files to temporal values for specific signals of
a bus-protocol of our choice. However, this would create an exploration barrier since 
bus-protocols require structured syntax, and most mutational fuzzers lack syntax 
awareness~\cite{afl_dictionary_blogpost}. In other words, the fuzzer would likely 
get stuck trying to synthesize a test file, that when mapped to spatio-temporal bus signal values,
produces a valid bus-transaction. Instead, we implement a harness that
decodes fuzzer-generated test files into sequences of properly structured bus 
transactions using a bus-centric grammar we describe below. Our current bus-centric harness
is implemented around the \ac{tlul} bus protocol~\cite{tilelink_spec} with a 32-bit data bus,
and illustrated in Fig.~\ref{fig:ot_tb_architecture}.

\subsubsection{\textbf{Bus-Centric Grammar}}
\label{subsubsection:hwf_bus_grammar}
\begin{figure}[t]
\centering
\includegraphics[width=0.4\textwidth]{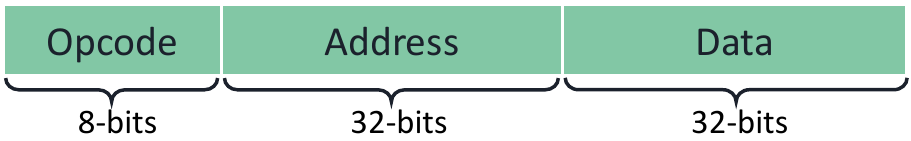}
\vspace*{-0.1in}
\caption{\footnotesize \textbf{\hwf{} Instruction.} 
A bus-centric harness (testbench) reads binary \textit{\hwf{} Instructions} from a fuzzer-generated test file, decodes them, and performs \ac{tlul} bus transactions to drive the \ac{dut} (Fig.\ref{fig:ot_tb_architecture}). Our \textit{\hwf{} Instructions} comprise a grammar (Tbl.~\ref{table:hwf_grammar}) that aid syntax-blind coverage-guided greybox fuzzers in generating valid bus-transactions to fuzz hardware.
}
\label{fig:hwf_instruction}
\figline{}
\vspace*{-0.1in}
\end{figure}

To translate fuzzer-generated test files into valid bus transactions we construct a
\hwf{} grammar. We format our grammar in a compact binary representation to facilitate integration
with popular greybox fuzzers that produce similar formats~\cite{afl,libfuzzer,honggfuzz}. To match
our bus-centric harness, we implement our grammar around the same \ac{tlul} bus 
protocol~\cite{tilelink_spec}. Our grammar consists of \textit{\hwf{} instructions} (Fig.~\ref{fig:hwf_instruction}), that contain: 1) an 8-bit opcode, 2) 32-bit address field, 
and 3) 32-bit data field. The opcode within each instruction determines the bus transaction
the harness performs. We describe the mappings between opcodes and \ac{tlul} bus transactions in Table~\ref{table:hwf_grammar}. 

Note, there are two properties of our grammar 
that leave room for various harness (testbench) implementations, which we study in 
\S\ref{subsection:hwf_grammar_eval}. First, while we define only three opcodes in 
our grammar, we represent the opcode with an entire byte, leaving it up to the 
harness to decide how to map \hwf{} opcode values to testbench 
actions. We do this for two reasons: 1) a byte is the smallest addressable unit in most 
software, facilitating the development of utilities to automate generating 
compact binary seed files (that comply with our grammar) from high-level markdown 
languages, and 2) choosing a larger opcode field enables adding more 
opcodes in the future, should we need to support additional operations in TileLink
bus protocol\cite{tilelink_spec}. Second, of the three opcodes we include, 
not all require address and data fields. Therefore, it is up to 
the harness to decide how it should process \hwf{} instructions. Should it read \textit{fixed} 
size instruction frames? Or \textit{variable} size instructions 
frames, depending on the opcode? To understand which interpretation of our \hwf{} 
grammar provides optimal constraints for greybox fuzzing, we study the 
performance of various binary encodings of our grammar in 
\S\ref{subsection:hwf_grammar_eval}.

\begin{table}[t]
\centering
\caption{\footnotesize \hwf{} Grammar.}
\label{table:hwf_grammar}
\begin{tabular}{lcll}
\hline
\hline
\textbf{Opcode} & \begin{tabular}[c]{@{}c@{}}\textbf{Address} \\ \textbf{Required}?\end{tabular} & \multicolumn{1}{c}{\begin{tabular}[c]{@{}c@{}}\textbf{Data} \\ \textbf{Required}?\end{tabular}} & \multicolumn{1}{c}{\textbf{Testbench Action}} \\
\hline
\hline
wait & no & no & advance the clock one period \\
\hline
read & yes & no & TL-UL Get (read) \\
\hline
write & yes & yes & TL-UL PutFullData (write) \\
\hline
\hline
\end{tabular}
\vspace*{-0.2in}
\end{table}

\section{\acl{hwfp}}\label{section:implementation}
\begin{figure}[t]
\centering
\includegraphics[width=0.3\textwidth]{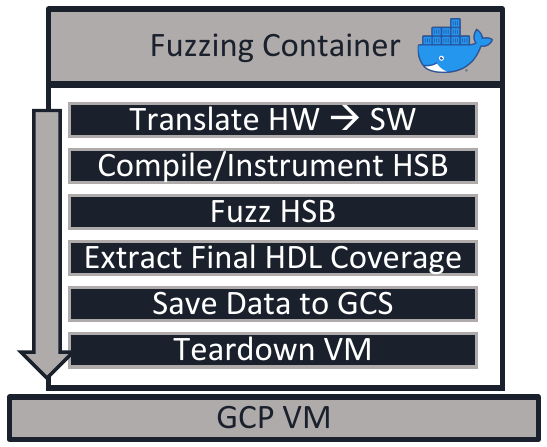}
\cprotect\caption{\footnotesize \textbf{\acf{hwfp}.} We design, implement, and open-source a \ac{hwfp} that is modeled after Google's OSS-Fuzz~\cite{oss_fuzz}. Our \ac{hwfp} enables us to verify \ac{rtl} hardware at scale using only open-source tools, a rarity in hardware \ac{dv}.
}
\label{fig:hwfp}
\figline{}
\vspace*{-0.2in}
\end{figure}
To fuzz hardware at scale we design, implement, and open-source~\cite{hwfp} a \acf{hwfp} modeled 
after Google's OSS-Fuzz (Fig.~\ref{fig:hwfp}). First, our pipeline builds a 
Docker image (from the Ubuntu 20.04 base image) containing a compiler (LLVM version 12.0.0), 
\ac{rtl} simulator (Verilator~\cite{verilator} version 4.0.4), software fuzzer, the target 
\ac{rtl} hardware, and a generic fuzzing harness 
(\S\ref{subsubsection:hwf_bus_harness}). From the image, a container is instantiated on a 
\ac{gcp} VM that:
\begin{enumerate}
    \item translates the \ac{dut}'s \ac{rtl} to a software model with 
      Verilator~\cite{verilator},
    \item compiles/instruments the \ac{dut} model, and links it with 
    the generic fuzzing harness (\S\ref{subsubsection:hwf_bus_harness}) and 
    simulation engine to create an \ac{hsb} (Fig.~\ref{fig:hw_sim_binary}),
    \item launches the fuzzer for a set period of time, using the \verb|timeout| 
      utility,
    \item traces final \ac{hdl} coverage of fuzzer-generated tests with 
      Verilator~\cite{verilator},
    \item saves fuzzing and coverage data to a \ac{gcs} bucket, and lastly
    \item tears down the VM.
\end{enumerate}
Note, for benchmarking, all containers are instantiated on their own 
\ac{gcp} \verb|n1-standard-2| VM with two vCPUs, 7.5\,GB of memory, 
50\,GB of disk, running Google's Container-Optimized OS.
In our current implementation, we use AFL~\cite{afl} (version 2.57b) as our fuzzer, 
but our \ac{hwfp} is designed to be fuzzer-agnostic. 

Unlike traditional hardware verification toolchains, our \ac{hwfp} uses \textit{only} 
open-source tools, allowing \ac{dv} engineers to save money on licenses, and spend 
it on compute. This not only enhances the deployability of our approach, but makes 
it ideal for adopting alongside existing hardware \ac{dv} workflows. This is 
important because rarely are new \ac{dv} approaches adopted without some overlap 
with prior (proven) techniques, since mistakes during hardware verification have 
costly repercussions.

\section{Evaluation - Part 1}\label{section:evaluation_p1}

In the first part of our evaluation, we address two technical questions
around fuzzing software models of \ac{rtl} hardware with software fuzzers.
First, \textit{how should
we interface coverage-guided software fuzzers with \acp{hsb}?} Unlike most software, 
\acp{hsb} contain other components---a testbench and simulation engine 
(Fig.~\ref{fig:hw_sim_binary})---that are \textit{not} the target of testing, 
yet the fuzzer must learn to manipulate in order to drive the \ac{dut}.
Second, \textit{how does \hwf{} compare with traditional dynamic verification methods, 
i.e., \ac{crv}, in terms of time to coverage convergence?}
To address this first set of questions, we perform several \ac{e2e} fuzzing analyses on over 
\numcircuits{} digital lock hardware designs with varying state-space complexities.

\subsection{Digital Lock Hardware}\label{subsection:lock_design}
\begin{figure}[t]
\centering
\includegraphics[width=0.49\textwidth]{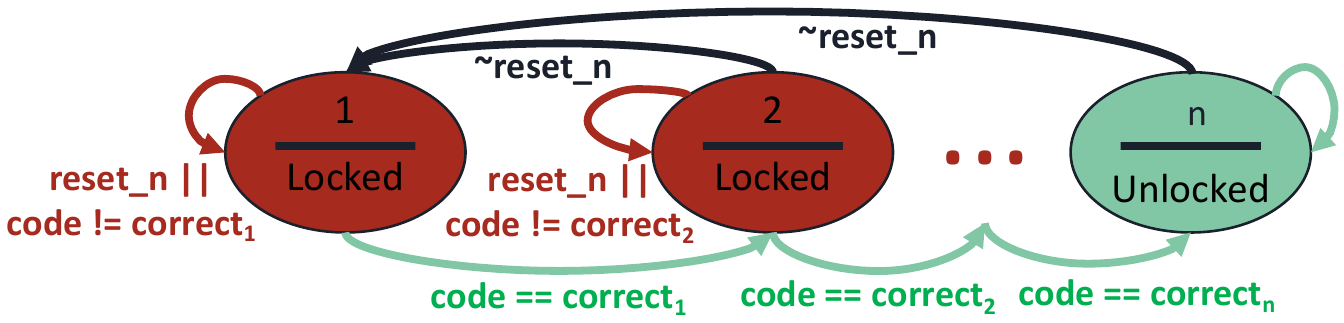}
\vspace*{-0.2in}
\caption{\footnotesize \textbf{Digital Lock \acs{fsm}.} We use a configurable digital lock 
(\ac{fsm} shown here) to demonstrate: 1) how to interface software fuzzers with hardware simulation binaries, and 2) the advantages of \hwf{} (vs.~traditional \ac{crv}). 
The digital lock \ac{fsm} can be configured in two dimensions: 1) total number
of states and 2) width (in bits) of input codes.
}
\label{fig:lock_fsm}
\figline{}
\vspace*{-0.2in}
\end{figure}

In this half of our evaluation, we fuzz various configurations of a digital 
lock, whose \ac{fsm} and \ac{hdl} are shown in Fig.~\ref{fig:lock_fsm} and 
List.~\ref{lst:lock_hdl}, respectively. We choose to study this design since the 
complexity of its state space is configurable, and therefore, ideal for stress testing 
various \ac{dv} methodologies. Specifically, the complexity is configurable in two 
dimensions: 1) the total number of states is configurable by tuning the size, $N$, 
of the single state register, and 2) the probability of choosing the correct 
unlocking code sequence is adjustable by altering the size, $M$, of the 
comparator/mux that checks input codes against hard-coded (random) values 
(List.~\ref{lst:lock_hdl}). We develop a utility in Rust, using the \verb|kaze| 
crate~\cite{kaze}, to auto-generate 480 different lock state machines of various 
complexities, i.e., different values of $N$, $M$, and random correct code 
sequences. 

\subsection{Digital Lock \acs{hsb} Architectures}\label{subsection:lock_tb}
\begin{figure}[t]
\centering
\includegraphics[width=0.49\textwidth]{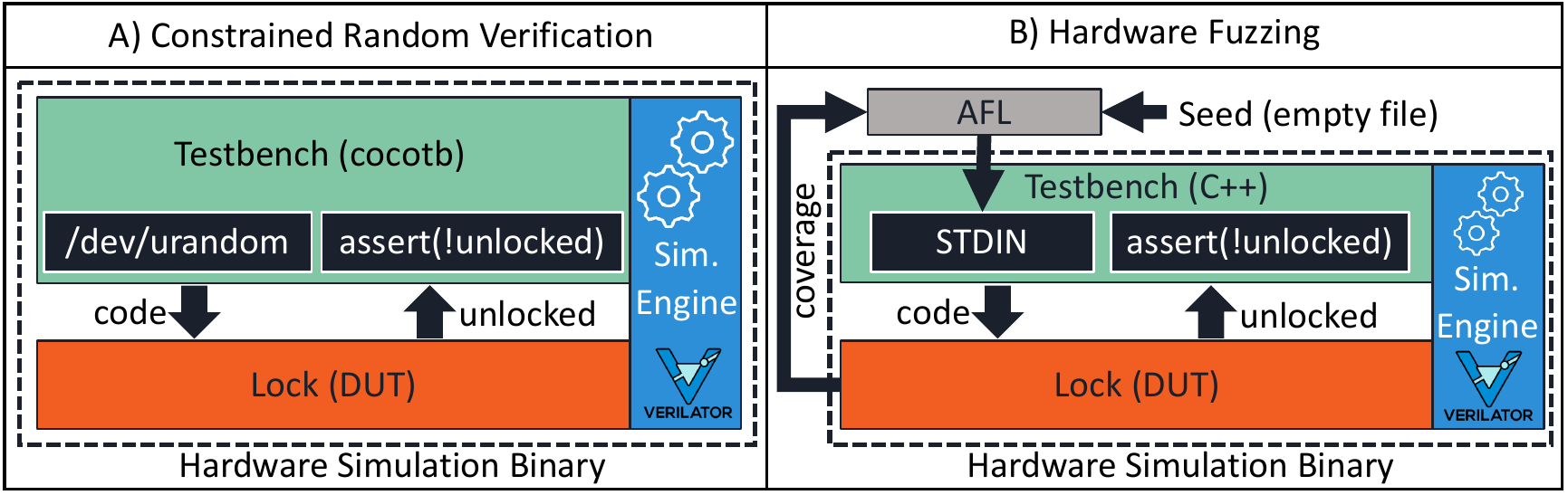}
\caption{\footnotesize \textbf{Digital Lock \ac{hsb} Architectures.}
(A) A traditional \ac{crv} architecture: random input code sequences 
are driven into the \ac{dut} until the unlocked state is 
reached. (B) A software fuzzer generates tests to drive the 
\ac{dut}. The fuzzer monitors coverage of the \ac{dut} during test execution and 
uses this information to generate future tests. Both \acp{hsb} are configured 
to terminate execution upon unlocking the lock using an \ac{sva} in the testbench
that signals the simulation engine (Fig.~\ref{fig:hw_sim_binary}) to abort.
}
\label{fig:lock_tb_architectures}
\figline{}
\vspace*{-0.2in}
\end{figure}

To study these designs, we construct two \ac{hsb} architectures 
(Fig.~\ref{fig:lock_tb_architectures}) using two hardware \ac{dv} methodologies: 
\ac{crv} and \hwf{}. The \ac{crv} architecture 
(Fig.~\ref{fig:lock_tb_architectures}A) attempts to unlock the lock through a 
brute-force approach, where random code sequences are driven into the \ac{dut} 
until the \textit{unlocked} state is reached. If the random sequence fails to 
unlock the lock, the \ac{dut} is reset, and a new random sequence is supplied. 
If the sequence succeeds, an \ac{sva} is violated, which terminates the 
simulation. The random code sequences are \textit{constrained} in the sense that only 
valid code sequences are driven into the \ac{dut}, i.e., 1) each code in the 
sequence is in the range $[0, 2^M)$ for locks with $M$-bit code comparators, 
and 2) sequences contain exactly $2^N-1$ input codes for locks with $2^N$ states. 
The \ac{crv} testbench is implemented with the cocotb~\cite{cocotb} framework 
and simulations are run with Verilator~\cite{verilator}.  

\begin{lstlisting}[
caption={SystemVerilog of Lock with $N$=$log2(\#\,states)$ and $M$-bit codes.
\vspace*{0.05in}
},
label={lst:lock_hdl}]
module lock(
  input reset_n,
  input clk,
  input [M-1:0] code,
  output unlocked
);
logic [N-1:0] state;
logic [M-1:0] correct_codes [N];

// Secret codes set to random values
for (genvar i = 0; i < N; i++) begin : secret_codes
  assign correct_codes[i] = <random value>;
end

assign unlocked = (state == '1) ? 1'b1 : 1'b0;

always @(posedge clk) begin
  if (!reset_n) begin
    state <= '0;
  end else if (!unlocked && code == correct_codes[state]) begin
    state <= state + 1'b1;
  end else begin
    state <= state;
  end
end
endmodule
\end{lstlisting}

Alternatively, the \hwf{} \ac{hsb} (Fig.~\ref{fig:lock_tb_architectures}B) 
takes input from a software fuzzer that generates code 
sequences for the \ac{dut}. The fuzzer initializes and 
checkpoints, a process running the \ac{hsb} (Fig.~\ref{fig:hw_sim_binary}), and 
repeatedly forks this process and tries various code sequence inputs. If an 
incorrect code sequence is supplied, the fuzzer forks a new process (equivalent to 
resetting the \ac{dut}) and tries again. If the correct code sequence is provided, 
an \ac{sva} is violated, which the fuzzer registers as a program crash. The difference between \ac{crv} and \hwf{} is that the fuzzer traces 
coverage during hardware simulation, and will \textit{save} past code sequences 
that get closer to unlocking the lock. These past sequences are then mutated to 
generate future sequences. Thus, past inputs are used to craft more 
\textit{intelligent} inputs in the future. To interface the software fuzzer with 
the \ac{hsb}, we:
\begin{enumerate}
    \item implement a C++ testbench harness from Algo.~\ref{algo:fuzzer_tb_harness} 
    that reads fuzzer-generated bytes from 
    \verb|stdin| and feeds them directly to the \verb|code| input of the lock.
    \item instrument the \ac{hsb} containing the \ac{dut} by 
    compiling it with \verb|afl-clang-fast++|.
\end{enumerate}

\subsection{Interfacing Software Fuzzers with Hardware}
\label{subsection:fuzzing_opts_eval}
There are two questions that arise when interfacing software fuzzers with \acp{hsb}. 
First, unlike most software applications, software models of hardware are not 
standalone binaries. They must be combined---typically by either static or dynamic 
linking---with a testbench and simulation engine to form an \ac{hsb} 
(\S\ref{subsubsection:background_hw_simulation}). Of these three components---\ac{dut}, 
testbench, and simulation engine---we seek to maximize coverage of \textit{only} the 
\ac{dut}. We do not want to waste fuzzing cycles on the testbench or simulation 
engine. Since coverage tracing instrumentation provides an indirect method to 
coarsely steer the fuzzer towards components of interest~\cite{bohme2017coverage}, 
it would be considered good practice to instrument just the \ac{dut} portion of 
the \ac{hsb}. However, while the \ac{dut} is ultimately what we want to fuzz, 
the fuzzer must learn to use the testbench and simulation engine to manipulate 
the \ac{dut}. Therefore, \textit{what components of the \ac{hsb} should we 
instrument to maximize fuzzer performance, yet ensure coverage convergence?} 

Second, when simulating hardware, the \ac{dut} must be reset to a clean state 
\textit{before} it can start processing inputs. Traditionally, the testbench portion 
of the \ac{hsb} performs this reset by asserting the \ac{dut}'s global reset signal 
for a set number of clock cycles. Since the fuzzer instantiates, and repeatedly forks 
the process executing the \ac{hsb}, this reset process will happen hundreds, or 
(potentially) thousands of times per second as each test execution is processed. 
While some software fuzzers~\cite{afl,libfuzzer} enable users to perform 
initialization operations \textit{before} the program under test is forked---meaning 
the \ac{dut} reset could be performed once, as each forking operation essentially 
sets the \ac{hsb} back to a clean state----this may not always the case. 
Moreover, it complicates fuzzer--\ac{hsb} integration, which contradicts the 
whole premise of our approach, i.e., low-overhead, design-agnostic 
\ac{cdg}. Therefore, we ask: \textit{is this fuzzing initialization feature 
\textit{required} to fuzz \acp{hsb}?}

\subsubsection{\textbf{Instrumenting \acp{hsb} for 
Fuzzing}}\label{subsubsection:instrumenting_hw_for_fuzzing}
To determine the components of the \ac{hsb} we should instrument, we measure the 
fuzzing run times to achieve approximate full \ac{fsm} coverage\footnote{We use the 
term \textit{approximate} when referring to \textit{full \ac{fsm} coverage}, since 
we are not excising the lock's reset state transitions (Fig.~\ref{fig:lock_fsm}) in 
these experiments.} of several lock designs, i.e., the time 
it takes the fuzzer to generate a sequence of input codes that \textit{unlocks each 
lock}. We measure this by modifying the fuzzer to terminate upon detecting the first
crash, which we produce using a single \ac{sva} that monitors the condition of the 
\textit{unlocked} signal (List.~\ref{lst:lock_hdl}). Specifically, using lock 
designs with 16, 32, and 64 states, and input codes widths of four bits, we 
construct \acp{hsb} following the architecture shown in 
Fig.~\ref{fig:lock_tb_architectures}B. For each \ac{hsb}, we vary the components we 
instrument by using different compiler settings for each component. First, we 
(na\"{i}vely) instrument \textbf{all} components, then only the \textbf{\ac{dut}}. 
Next, we fuzz each \ac{hsb} 50 times, seeding the fuzzer with an empty file in each 
experiment.

\begin{figure}[t]
\centering
\includegraphics[width=0.49\textwidth]{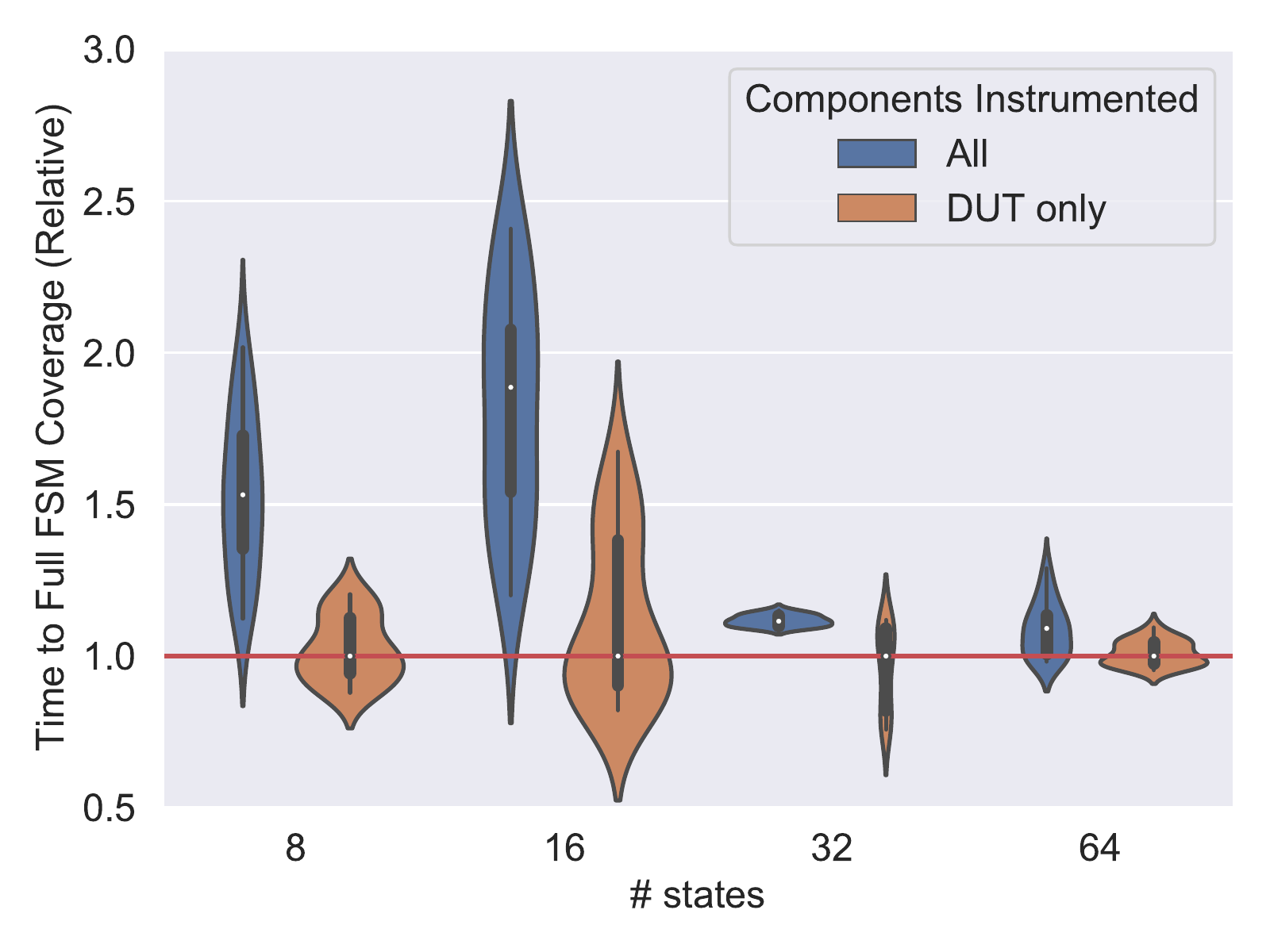}
\vspace*{-0.3in}
\cprotect\caption{\footnotesize \textbf{Instrumentation Level vs. Coverage 
Convergence Rate.} Distribution of fuzzer run times required to \textit{unlock} various sized digital locks (code widths are fixed at four bits), i.e., achieve $\approx$ full \ac{fsm} coverage. For each \ac{hsb}, we vary the components we instrument for coverage tracing. Run times are normalized to the median \ac{dut}-only instrumentation level (orange) across each lock size (red line). While the fuzzer uses the testbench and simulation engine to manipulate the \ac{dut}, instrumenting only the \ac{dut} \textit{does not} hinder the coverage convergence rate of the fuzzer. Rather, it improves it when \ac{dut} sizes are small, compared to the simulation engine and testbench (Fig.~\ref{fig:hwf_components_bbs}).
}
\label{fig:hwf_instrumentation_levels_eval}
\figline{}
\vspace*{-0.15in}
\end{figure}

\begin{figure}[t]
\centering
\includegraphics[width=0.45\textwidth]{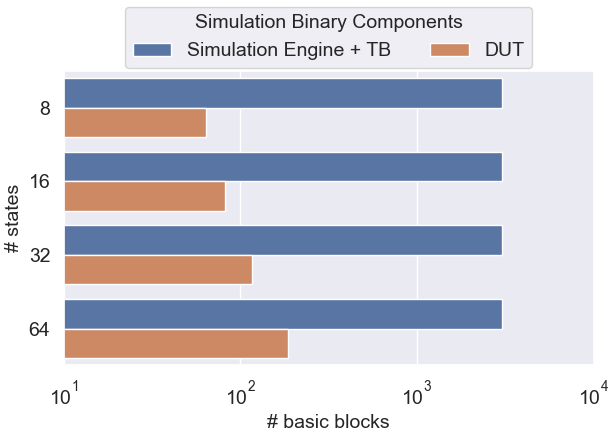}
\vspace*{-0.15in}
\cprotect\caption{\footnotesize \textbf{Basic Blocks per Simulation Binary Component.}
We break down the number of basic blocks that comprise the three components within 
\acp{hsb} of different size locks (Fig.~\ref{fig:lock_fsm} \& 
List.~\ref{lst:lock_hdl}), generated by Verilator~\cite{verilator}: simulation engine and testbench (TB), and \ac{dut}. As locks increase in size, defined by the number of 
\ac{fsm} states (code widths are fixed to 4 bits), so do the number of basic blocks in
their software model.
}
\label{fig:hwf_components_bbs}
\figline{}
\vspace*{-0.15in}
\end{figure}

We plot the distribution of fuzzing run times in 
Fig.~\ref{fig:hwf_instrumentation_levels_eval}. Since fuzzing is an inherently 
random process, we plot only the middle third of run times across all 
instrumentation levels and lock sizes. Moreover, all run times are normalized to 
the median \ac{dut}-only instrumentation run times (orange) across each lock size. 
In addition to plotting fuzzing run times, we plot the number of basic blocks 
within each component of the \ac{hsb} in Fig.~\ref{fig:hwf_components_bbs}. 
Across all lock sizes, we observe that only instrumenting the \ac{dut} does 
not handicap the fuzzer, but rather \textit{improves the rate of coverage 
convergence}! In fact, we perform a Mann-Whitney U test, with a 0.05 significance 
level, and find all the run-time improvements to be statistically significant. 
Moreover, we observe that even though the run-time improvements are less 
significant as the \ac{dut} size increases compared to the simulation engine 
and testbench (Fig.~\ref{fig:hwf_components_bbs}), instrumenting only the 
\ac{dut} never handicaps the fuzzer performance.

\llbox{
\textbf{Key Insight:} Instrumenting only the \ac{dut} portion of the 
\ac{hsb} does not impair the fuzzer's ability to drive the \ac{dut}, rather, 
it improves fuzzing speed.
}


\subsubsection{\textbf{Hardware Resets vs.~Fuzzer Performance}}
\label{subsubsection:hw_resets_opt}

To determine if \ac{dut} resets present a performance bottleneck, we measure the 
degradation in fuzzing performance due to the repeated simulation of \ac{dut} resets. 
We take advantage of a unique feature of a popular greybox fuzzer~\cite{afl} that 
enables configuring the exact location of initializing the 
\textit{fork server}.\footnote{By default, AFL~\cite{afl} instantiates a 
process from the binary under test, pauses it, and repeatedly forks it to 
create identical processes to feed test inputs to. The component 
of AFL that performs process forking is known as the \textit{fork server}.} 
This enables the fuzzer to perform any 
program-specific initialization operations \textit{once}, prior to forking children 
processes to fuzz. Using this feature, we repeat the same fuzzing run time analysis 
performed in \S\ref{subsubsection:instrumenting_hw_for_fuzzing}, except we instrument 
all simulation binary components, and compare two variations of the digital lock 
\ac{hsb} shown in Fig.~\ref{fig:lock_tb_architectures}B. In one testbench, we use 
the default fork server initialization location: at the start of \verb|main()|. 
In the other testbench, we initialize the fork server \textit{after} the point 
where the \ac{dut} has been reset.

\begin{figure}[t]
\centering
\includegraphics[width=0.45\textwidth]{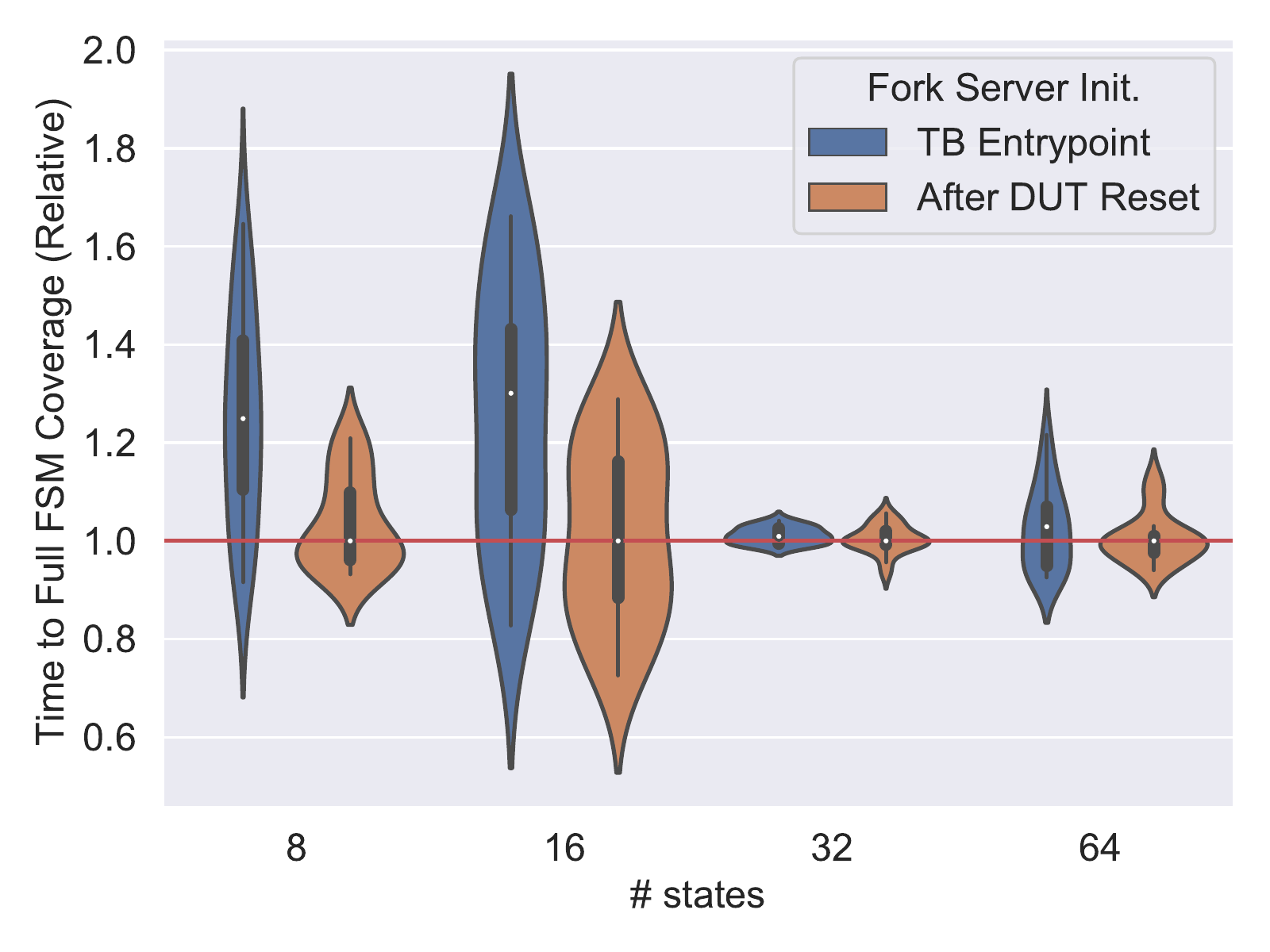}
\vspace*{-0.2in}
\cprotect\caption{\footnotesize \textbf{Hardware Resets vs. Fuzzer Performance.} Fuzzing run times across across digital locks (similar to Fig.~\ref{fig:hwf_instrumentation_levels_eval}) with different fork server initialization locations in the testbench to eliminate overhead due to the repeated simulation of hardware \ac{dut} resets. \ac{dut} resets are only a fuzzing bottleneck when \acp{dut} are small, reducing fuzzer--\ac{hsb} integration complexity.
}
\label{fig:hwf_fs_opt}
\figline{}
\vspace*{-0.2in}
\end{figure}
\begin{figure*}[t]
\centering
\includegraphics[width=0.9\textwidth]{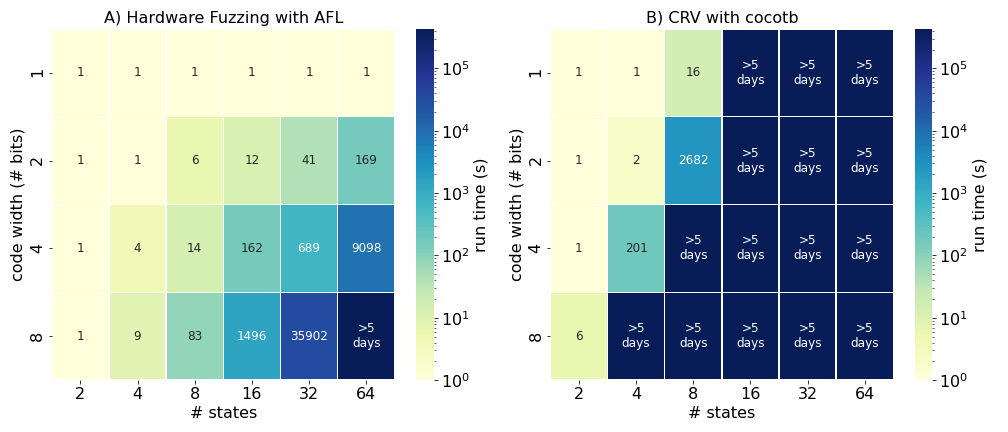}
\vspace*{-0.15in}
\cprotect\caption{\footnotesize \textbf{Hardware Fuzzing vs. \acs{crv}.}
Run times for both \hwf{} (A) and \ac{crv} (B) to achieve $\approx$\,full \ac{fsm} coverage of various digital lock (Fig.~\ref{fig:lock_fsm}) designs---i.e., time to unlock the lock---using the testbench architectures shown in Fig.~\ref{fig:lock_tb_architectures}. 
Run times are averaged across 20 trials for each lock design---defined by a 
(\# states, code width) pair---and \ac{dv} method combination. Across these 
designs, \hwf{} achieves full \ac{fsm} coverage faster than 
traditional \ac{crv} approaches, by over two orders of magnitude.
}
\label{fig:fuzzing_vs_crv_eval}
\figline{}
\vspace*{-0.15in}
\end{figure*}

Fig.~\ref{fig:hwf_fs_opt} shows our results. Again, we drop outliers by plotting only 
the middle third of run times across all lock sizes and fork server initialization 
points. Additionally, we normalize all run times to the median ``after \ac{dut} 
reset'' run times (orange) across each lock size. From these results, we 
apply the Mann-Whitney U test (with 0.05 significance level) between run times. 
This time, only locks with 8 and 16 states yield p-values less than 0.05. This 
indicates the overhead of continuously resetting the \ac{dut} during fuzzing diminishes as 
the \ac{dut} increases in complexity. Additionally, we note that even the largest 
digital locks we study (64 states), are smaller than the smallest OpenTitan core, 
the RISC-V Timer, in terms of number of basic blocks in the software model 
(Fig.~\ref{fig:hwf_components_bbs} \& Table~\ref{table:ip_block_complexity}).

\llbox{
\textbf{Key Insight:} Overhead from simulating hardware resets while fuzzing 
is minimal, especially in large designs, further reducing fuzzer--\ac{hsb} 
integration efforts.
}

\subsection{Hardware Fuzzing vs.~\ac{crv}}
\label{subsection:fuzzing_vs_crv_eval}
Using the techniques we learned from above, we perform a run-time comparison 
analysis between \hwf{} and \ac{crv},\footnote{\ac{crv} 
is widely deployed in any \ac{dv} testbenches built around the cocotb~\cite{cocotb} 
or \ac{uvm}~\cite{uvm} frameworks, e.g., all OpenTitan~\cite{opentitan} \ac{ip} 
core testbenches.} the current 
state-of-the-art hardware dynamic verification technique.
We perform these experiments using 
digital locks of various complexities, from 2 to 64 states, and code widths 
of 1 to 8 bits. The two \ac{hsb} architectures 
we compare are shown in Fig.~\ref{fig:lock_tb_architectures}, and discussed 
in \S\ref{subsection:lock_tb}. Note, the fuzzer was again seeded with 
an empty file to align its starting state with the \ac{crv} tests.

Similar to our instrumentation and reset experiments 
(\S\ref{subsection:fuzzing_opts_eval}) we measure the fuzzing \textit{run times} 
required to achieve $\approx$ full \ac{fsm} coverage of each lock design, 
i.e., the time to \textit{unlock each lock}.
We illustrate these run times in heatmaps shown in 
Fig.~\ref{fig:fuzzing_vs_crv_eval}. 
We perform 20 trials for each experiment and average these run times in each square 
of a heatmap. While the difference between the two approaches is indistinguishable 
for extremely small designs, the advantages of \hwf{}
become apparent as designs increase in complexity. For medium to larger lock 
designs, \hwf{} achieves full \ac{fsm} coverage faster than 
\ac{crv} by over two orders-of-magnitude, even when the fuzzer is seeded with an 
empty file. Moreover, many \ac{crv} experiments were terminated early 
(after running for five days) to save money on \ac{gcp} instances.

\llbox{
\textbf{Key Insight:} 
\hwf{} is a low-cost, low-overhead \ac{cdg} approach for hardware \ac{dv}.
}

\section{Evaluation - Part 2}\label{section:evaluation_p2}
In the second part of our evaluation, we address two remaining questions. First, 
\textit{how should we format our grammar to enable the fuzzer to learn it quickly?} 
To facilitate widespread deployment of \hwf{}, it is imperative 
\ac{dv} engineers do \textit{not} have to tailor fuzzing harnesses (testbenches)
to specific designs, as is the case with existing \ac{cdg} methods~\cite{fine2003coverage,cieplucha2019metric,ioannides2011introducing,teplitsky2015coverage,guzey2007coverage,wang2018accelerating,laeufer2018rfuzz}. 
Lastly, \textit{how does \hwf{} perform in practice on real hardware \ac{ip} cores?}
To address these questions, we perform \ac{e2e} fuzzing analyses on
four commercial hardware cores from Google's OpenTitan~\cite{opentitan} \ac{soc}.

\subsection{OpenTitan \acs{ip}}\label{subsubsection:opentitan_cores}
\begin{figure}[t]
\centering
\includegraphics[width=0.4\textwidth]{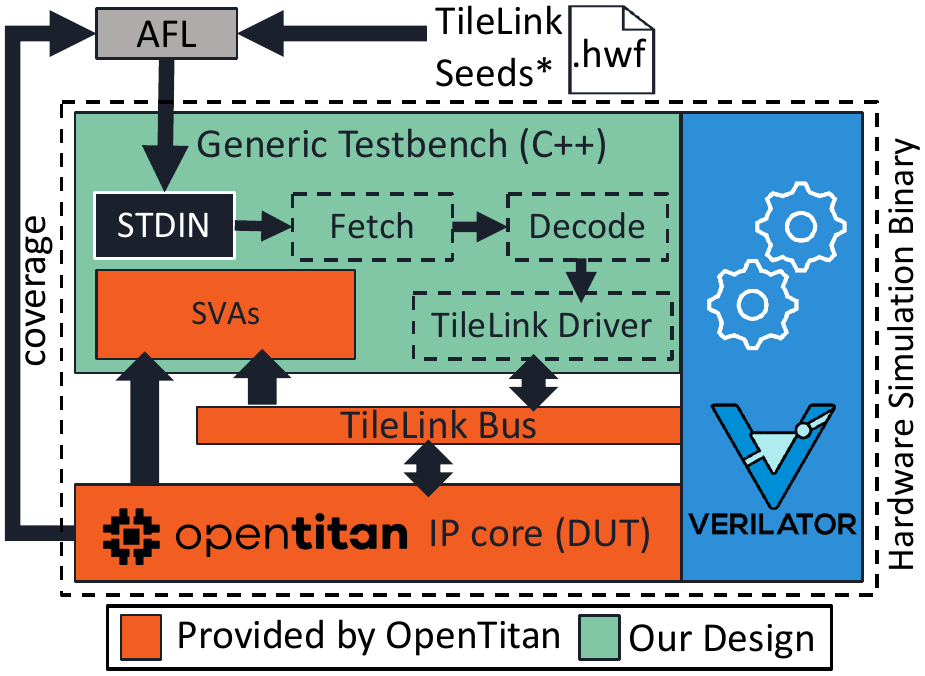}
\cprotect\caption{\footnotesize \textbf{OpenTitan \ac{hsb} 
Architecture.} A software fuzzer learns to generate 
fuzzing \textit{instructions} (Fig.~\ref{fig:hwf_instruction})---from \verb|.hwf| seed files---based on a hardware fuzzing grammar (\S\ref{subsubsection:hwf_bus_grammar}). It pipes these instructions to \verb|stdin| where a generic C++ fuzzing harness fetches/decodes them, and performs the corresponding TileLink bus operations to drive the \ac{dut}. \acp{sva} are evaluated during execution of the \ac{hsb}, and produce a program crash (if violated), that is caught and reported by the software fuzzer.
}
\label{fig:ot_tb_architecture}
\figline{}
\vspace*{-0.2in}
\end{figure}

The four OpenTitan \ac{ip} blocks we study are the: AES, HMAC, KMAC, and RISC-V 
Timer cores. While each core performs different functions, they all conform to the 
OpenTitan \textit{Comportability Specification}~\cite{ot_comportability_guide}, 
which implies \textbf{they are all controlled via reads and writes to 
memory-mapped registers over a \ac{tlul} bus}. By adhering to a uniform bus 
protocol, we are able to re-use a generic fuzzing harness 
(Fig.~\ref{fig:ot_tb_architecture}), facilitating the deployability of our 
approach. Below, we highlight the functionality of each \ac{ip} core. 
Additionally, in Table~\ref{table:ip_block_complexity}, we report the 
complexity of each \ac{ip} core in both the hardware and software domains, 
in terms of \ac{loc}, number of basic blocks, and number of \acp{sva} provided 
in each core's \ac{hdl}. Software models of each hardware design are produced 
using Verilator, as we describe in \S\ref{subsection:translating_hw_to_sw}.

\subsubsection{\textbf{AES}}
The OpenTitan AES core implements the Advanced Encryption 
Standard with key sizes of 128, 192, and 256 bits, and with the following 
cipher block modes: ECB, CBC, CFB, OFB, and CTR. Configuration settings, keys, 
and plaintext are delivered to the core through TileLink write operations to 
memory-mapped registers in a documented address range. Likewise, ciphertext is 
retrieved from the core through TileLink read operations. The core targets 
medium performance (one clock cycle per round of encryption). It implements a 
128-bit wide data path---shared by encryption and decryption operations---that 
translates to encryption/decryption latencies of 12, 14, and 16 clock cycles 
per 128-bit plaintext block, in 128, 192, and 256 bit key modes, respectively. 
Of the cores we study, it is the second most complex in terms of \ac{loc} in 
both the hardware (\ac{hdl}) and software domains 
(Table~\ref{table:ip_block_complexity}).

\subsubsection{\textbf{HMAC}}
The OpenTitan HMAC implements a SHA-256 hash message 
authentication code generator for the purpose of checking the integrity of 
incoming messages. The HMAC core can operate in two modes: 1) SHA-256 mode 
only, or 2) HMAC mode. In the former mode, the core simply computes the 
SHA-256 hash of a provided message. In the latter mode, the core computes the 
HMAC (defined in RFC 2104~\cite{rfc2104}) of a message using the SHA-256 
hashing algorithm and a provided secret key. Regardless of mode, the SHA-256 
engine operates on 512-bit message chunks at any given time, provided to the 
core through a message FIFO. Input messages can be read little- or big-endian 
and likewise, message digests can be stored in output registers either 
little- or big-endian. Configuration settings, input messages, HMAC keys, and 
operation commands are delivered to the core through TileLink write operations 
to memory-mapped registers. Likewise, message digests are retrieved from the 
core through TileLink read operations. In its current state, the core can 
hash a single 512-bit message in 80 clock cycles, and can compute its 
HMAC in 340 clock cycles. Of the cores we study, it is approximately half as 
complex as the AES core, in terms of \ac{loc} in both the hardware and 
software domains (Table~\ref{table:ip_block_complexity}).

\subsubsection{\textbf{KMAC}}
The OpenTitan KMAC core is similar to the HMAC core, except it implements a 
Keccak Message Authentication Code~\cite{kelsey2016sha} and SHA-3 hashing 
algorithms. However, compared to the HMAC core, the KMAC core is more complex, 
as there are several more configurations. Specifically, there are many SHA-3 
hashing functions that are supported---SHA3-224/256/384/512, SHAKE128/256, and 
cSHAKE128/256---and the $Keccak-f$ function (by default) operates on 1600 bits of 
internal state. Like the HMAC core, the KMAC core can simply compute hashes or 
message authentication codes depending on operation mode, and input messages/output 
digests can be configured to be read/stored in little- or big-endian. The time 
to process a single input message block is dominated by computing the $Keccak-f$ 
function, which takes 72 clock cycles for 1600 bits of internal state, in the 
current implementation of the core. Configuration settings, input messages, 
output digests, keys, and operation commands are all communicated to/from the 
core through TileLink writes/reads to memory-mapped registers.

Of the cores we study, the KMAC core is the most complex, especially in the 
software domain (Table~\ref{table:ip_block_complexity}). The software model of the
KMAC core contains almost 120k lines of C++ code. This is mostly an artifact of 
how Verilator maps dependencies between large registers and vectored signals: 
it creates large multidimensional arrays 
and maps each corresponding index at the word granularity. Fortunately, this 
artifact is optimized away during compilation, and the number of basic blocks 
in the \ac{dut} portion of the \ac{hsb} is reduced. 

\begin{table}[!t]
\centering
\caption{\footnotesize OpenTitan \acs{ip} Core Complexity in HW and SW Domains.
}
\label{table:ip_block_complexity}
\begin{threeparttable}
\begin{tabular}{l c c c c}
\hline
\hline
\footnotesize \textbf{\acs{ip} Core} &
\footnotesize \textbf{HW \acs{loc}} &
\footnotesize \textbf{SW \acs{loc}} &
\footnotesize \textbf{\# Basic Blocks}\textasteriskcentered &
\footnotesize \textbf{\# \acsp{sva}}\textdagger \\
\hline
\hline
\footnotesize AES & 
\footnotesize 4,562 & 
\footnotesize 38,036 & 
\footnotesize 3,414 &
\footnotesize 53 \\
\hline
\footnotesize HMAC & 
\footnotesize 2,695 & 
\footnotesize 18,005 & 
\footnotesize 1,764 &
\footnotesize 30 \\
\hline
\footnotesize KMAC & 
\footnotesize 4,585 & 
\footnotesize 119,297 & 
\footnotesize 6,996 &
\footnotesize 44 \\
\hline
\footnotesize RV Timer & 
\footnotesize 677 & 
\footnotesize 3,111 & 
\footnotesize 290 &
\footnotesize 8 \\
\hline
\hline
\end{tabular}
\begin{tablenotes}
    \item[\textasteriskcentered] \footnotesize \# of basic blocks in compiled software model with O3 optimization.
    \item[\textdagger] \footnotesize \# of \aclp{sva} included in \ac{ip} \ac{hdl} at time of writing.
\end{tablenotes}
\vspace*{-0.15in}
\end{threeparttable}
\end{table}

\subsubsection{\textbf{RV-Timer}}
The OpenTitan RISC-V timer core is the simplest core we fuzz. It consists of a 
single 64-bit timer with 12-bit prescaler and an 8-bit step configurations. 
It can also generate system interrupts upon reaching a pre-configured time 
value. Like the other OpenTitan cores, the RV-Timer core is configured, 
activated, and deactivated via TileLink writes to memory-mapped registers.

\begin{figure*}[t]
\centering
\includegraphics[width=\textwidth]{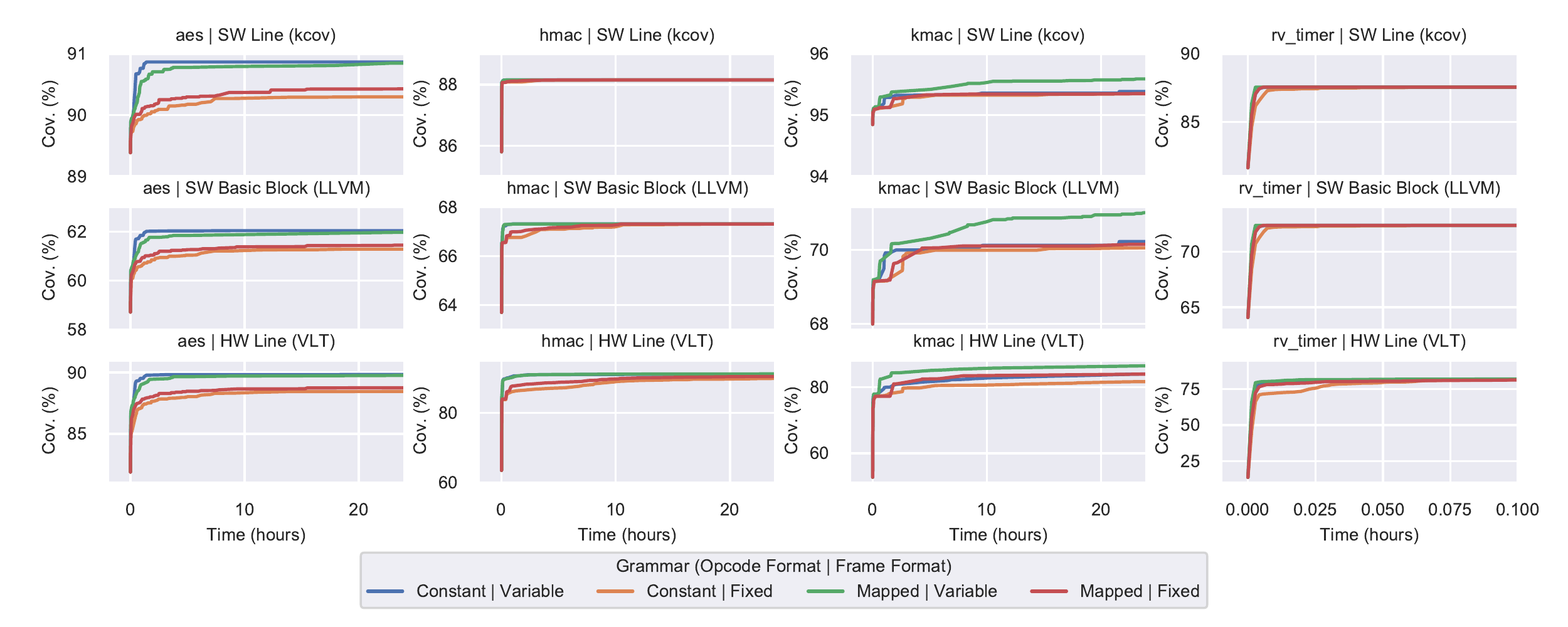}
\vspace*{-0.3in}
\cprotect\caption{\footnotesize \textbf{Coverage Convergence vs. Hardware Fuzzing Grammar.}
Various software and hardware coverage metrics over fuzzing time across four OpenTitan~\cite{opentitan} \ac{ip} cores and hardware fuzzing grammar variations (\S\ref{subsection:hwf_grammar_eval}). In the first row, we plot \textit{line coverage} of the software models of each hardware core computed using \verb|kcov|. In the second row, we plot \textit{basic block coverage} computed using \verb|LLVM|. In last row, we plot \ac{hdl} line coverage (of the hardware itself) computed using \verb|Verilator|~\cite{verilator}. From these results we formulate two conclusions: 1) coverage in the software domain correlates to coverage in the hardware domain, and 2) the \hwf{} grammar with \textit{variable} instruction frames is best for greybox fuzzers that prioritize small test files.
}
\label{fig:hwf_grammar_opt}
\figline{}
\vspace*{-0.2in}
\end{figure*}

\subsection{Optimizing the Hardware Fuzzing Grammar}
\label{subsection:hwf_grammar_eval}

Recall, to facilitate widespread adoption of \hwf{} we design a 
generic testbench fuzzing harness that decodes a grammar and performs
corresponding \ac{tlul} bus transactions to exercise the \ac{dut} 
(Fig.~\ref{fig:ot_tb_architecture}). However, there are implementation
questions surrounding how the grammar should be decoded 
(\S\ref{subsubsection:hwf_bus_grammar}):
\begin{enumerate}
    \item \textit{How should we decode 8-bit opcodes when the opcode space 
      defines less than $2^8$ valid testbench actions?}
    \item \textit{How should we pack \hwf{} instruction \textit{frames} that 
      conform to our grammar?}
\end{enumerate}

\subsubsection{\textbf{Opcode Formats}}
\label{subsubsection:opcode_format_eval}
In its current state, we define three opcodes in our grammar that correspond to 
three actions our generic testbench can perform (Table \ref{table:hwf_grammar}): 
1) \textbf{wait} one clock cycle, 2) \ac{tlul} \textbf{read}, and 
3) \ac{tlul} \textbf{write}. However, we chose to 
represent these opcodes with a single byte (Fig.~\ref{fig:hwf_instruction}). 
Choosing a larger field than necessary has 
implications regarding the fuzzability of our grammar. In its current state, 253 of 
the 256 possible opcode values may be useless depending on how they are decoded by 
the testbench. Therefore we propose, and empirically study, two design choices for 
decoding \hwf{} opcodes into testbench actions:
\begin{itemize}
    \item \textbf{Constant}: \textit{constant} values are used to represent each 
      opcode corresponding to a single testbench action. Remaining opcode values 
      are decoded as \textit{invalid}, and ignored.
    \item \textbf{Mapped}: equal sized ranges of opcode values are \textit{mapped} 
      to valid testbench actions. No invalid opcode values exist.
\end{itemize}

\subsubsection{\textbf{Instruction Frame Formats}}
\label{subsubsection:frame_format_eval}

Of the three actions our testbench can perform---wait, read, and write---some require 
additional information. Namely, the \ac{tlul} read action requires a 32-bit address 
field, and the \ac{tlul} write action requires 32-bit data and address fields. 
Given this, there are two natural ways to decode \hwf{} instructions 
(Fig.~\ref{fig:hwf_instruction}):
\begin{itemize}
    \item \textbf{Fixed}: a \textit{fixed} instruction frame size is decoded regardless
    of the opcode. Address and data fields could go unused depending on the opcode.
    \item \textbf{Variable}: a \textit{variable} instruction frame size is decoded. 
      Address and data fields are only appended to opcodes that correspond to 
      \ac{tlul} read and write testbench actions. No address/data information 
      goes unused.
\end{itemize}

\subsubsection{\textbf{Results}}
\label{subsubsection:hwf_grammar_eval_results}

To determine the optimal \hwf{} grammar, we fuzz four OpenTitan \ac{ip} blocks---the 
AES, HMAC, KMAC, and RV-Timer---for 24 hours
using all combinations of opcode and instruction frame formats mentioned above. 
For each core we seed the fuzzer with 8--12 binary \hwf{} seed files 
(in the corresponding \hwf{} grammar) that correctly drive each core, with the 
exception of the RV-Timer core, which we seed with a single wait operation 
instruction due to its simplicity. For each experiment, we extract and plot three 
\ac{dut} coverage metrics over fuzz times in Fig.~\ref{fig:hwf_grammar_opt}. 
These metrics include: 1) line coverage of the \ac{dut} software model, 2) basic 
block coverage of the same, and 3) line coverage of the \ac{dut}'s \ac{hdl}. 
Software line coverage is computed using \verb|kcov|~\cite{kcov}, software basic 
block coverage is computed using \verb|LLVM|~\cite{llvm}, and hardware line coverage 
is computed using \verb|Verilator|~\cite{verilator}. Since we perform 10 repetitions 
of each fuzzing experiment, we average and consolidate each coverage time series 
into a single trace. 

From these results we draw two conclusions. First, \textit{variable} 
instruction frames seem to perform better than fixed frames, especially early
in the fuzzing exploration. Since AFL prioritizes keeping test files small,
we expect variable sized instruction
frames to produce better results, since this translates to longer
hardware test sequences, and therefore deeper possible explorations of the
(sequential) state space. Second, the opcode type seems to make little difference,
for most experiments, since there are only 256 possible values, 
a search space AFL can explore very quickly.
Lastly, we point out that for simple cores, like the RV-Timer, \hwf{} is able to 
achieve $\approx$85\% \ac{hdl} line coverage in less than a minute
(hence we do not plot the full 24-hour trace).

\llbox{
\textbf{Key Insights:} 
\begin{enumerate}
    \item \hwf{} instructions with \textbf{variable} frames are optimal for fuzzers that prioritize small input files, therefore resulting in longer temporal test \textit{sequences}.
    \item Increasing coverage in the software domain, translates to 
    the hardware domain.
\end{enumerate}
}

\subsection{\hwf{} in Practice}\label{subsection:hwf_in_practice}

Finally, we address the question: \textit{How does \hwf{} perform in practice?} 
First, we show that with no knowledge of how to properly use the \ac{dut}, we 
achieve almost 90\% \ac{hdl} line coverage across the OpenTitan~\cite{opentitan} 
cores we study. Second, we compare \hwf{} against the most popular \ac{dv} 
technique today, \ac{crv}, demonstrating over two orders-of-magnitude faster 
coverage-convergence times.

\begin{figure}[t]
\centering
\includegraphics[width=0.48\textwidth]{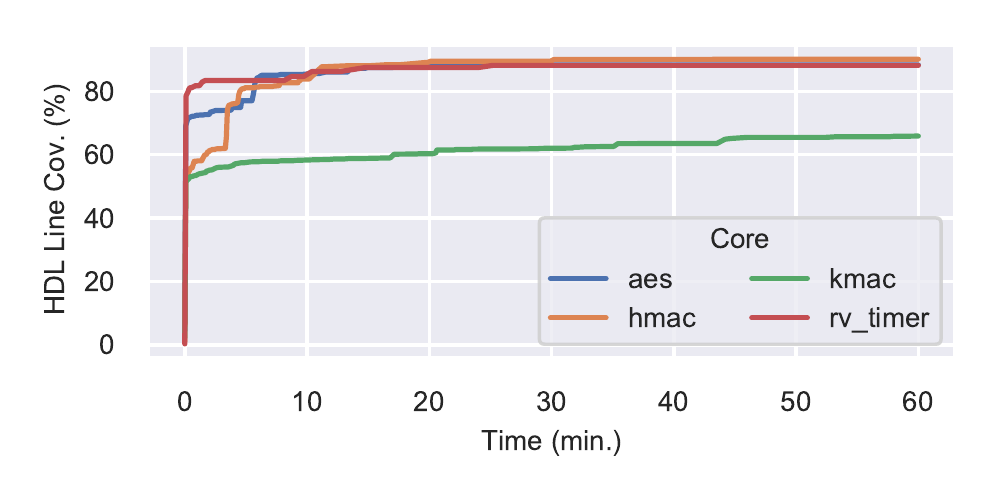}
\vspace*{-0.2in}
\cprotect\caption{\footnotesize \textbf{Coverage vs. Time Fuzzing with Empty Seeds.} Fuzzing four OpenTitan~\cite{opentitan} \ac{ip} cores for one hour, seeding the fuzzer with an empty file in each case, yields over 88\% \ac{hdl} line coverage in three out of four designs.
}
\label{fig:hwf_no_seeds}
\figline{}
\vspace*{-0.2in}
\end{figure}

Unlike most software applications that are fuzzed~\cite{oss_fuzz}, we observe that 
software models of hardware are quite small (Table~\ref{table:ip_block_complexity}). 
So, we decided to experiment fuzzing each OpenTitan core we study for one hour, 
using a single empty seed file as starting input. We plot the results of this 
experiment in Fig.~\ref{fig:hwf_no_seeds}. After only one hour of fuzzing with no 
proper starting seeds, we achieve over 88\% \ac{hdl} line coverage across three of 
the four OpenTitan \ac{ip} cores we study, and over 65\% coverage of the 
remaining design.

\section{Discussion}\label{section:discussion}
\subsubsection{\textbf{Detecting Bugs During Fuzzing}}
The focus of \hwf{} is to provide a scalable yet flexible solution for integrating 
\ac{cdg} with hardware simulation. However, \textit{test generation} and 
\textit{hardware simulation} comprise only two-thirds of the hardware verification 
process (\S\ref{subsection:background_dynamic_verification}). The final, and 
arguably most important, step is detecting incorrect hardware behavior, i.e., 
\textit{test evaluation} in \S\ref{subsubsection:background_test_evaluation}. 
For this there are two approaches: 1) invariant checking and 2) (gold) model checking. 
In both cases, we trigger \ac{hsb} crashes upon detecting incorrect hardware behavior,
which software fuzzers log. For invariant checks, we use \acp{sva} that send the \ac{hsb} 
process the \verb|SIGABRT| signal upon assertion violation. Likewise, for gold model 
checking testbenches any mismatches between models results in a \verb|SIGABRT|. 

\subsubsection{\textbf{Additional Bus Protocols}} To provide a design-agnostic
interface to fuzz \ac{rtl} hardware, we develop a design-agnostic testbench harness 
(Fig.~\ref{fig:ot_tb_architecture}). Our harness decodes fuzzer-generated 
tests using a bus-specific grammar (\S\ref{subsubsection:hwf_bus_grammar}), and 
produces corresponding \ac{tlul} bus transactions that drive a \ac{dut}. In our 
current implementation, our generic testbench harness conforms to the \ac{tlul} bus 
protocol~\cite{tilelink_spec}. As a result, we can fuzz any \ac{ip} core that speaks 
the same bus protocol (e.g., all OpenTitan cores \cite{opentitan}). To fuzz 
cores that speak other bus protocols (e.g., Wishbone, AMBA, Avalon, etc.), 
users can simply write a new harness for the bus they wish to support.

\subsubsection{\textbf{Hardware without a Bus Interface}}
For hardware cores that perform I/O over a generic set of ports that do not
conform to any bus protocol, we provide a generic testbench harness that maps 
fuzzer-generated input files across spatial and temporal domains by interpreting 
each fuzzer-generated file as a sequence of \ac{dut} inputs 
(Algo.~\ref{algo:fuzzer_tb_harness}). We demonstrate this \hwf{} configuration 
when fuzzing various digital locks (Fig.~\ref{fig:lock_tb_architectures}B). 
However, if inputs require any structural dependencies, we advise developing 
a grammar and corresponding testbench---similar to our bus-specific 
grammar (\S\ref{subsubsection:hwf_bus_grammar})---to aid the fuzzer in 
generating valid test cases. Designers can use the lessons in this paper to 
guide their core-specific grammar designs.

\subsubsection{\textbf{Limitations}}
While \hwf{} is both efficient and design-agnostic, there are some limitations. 
First, unlike software there is no notion of a \textit{hardware sanitizer}, 
that can add safeguards against generic classes of hardware bugs for the 
fuzzer to sniff out. While we envision hardware sanitizers being a future 
active research area, for now, \ac{dv} engineers must create invariants
or gold models to check design behavior against for the fuzzer to find 
crashing inputs.
Second, there is notion of analog behavior in \ac{rtl} hardware,
let along in translated software models. In its current implementation,
\hwf{} is not effective against detecting side-channel vulnerabilities 
that rely on information transmission/leakage through analog domains.

\section{Related Work}\label{section:related_work}
There are two categories of prior \ac{cdg} approaches: 1) design-agnostic and 2) 
design-specific.

\subsubsection{\textbf{Design-Agnostic}} 
Laeufer \etal{}'s \verb|RFUZZ|~\cite{laeufer2018rfuzz} is the most relevant prior work, 
which attempts to build a full-fledged design-agnostic \ac{rtl} fuzzer. To achieve 
their goal, they propose a new \ac{rtl} coverage metric---\textit{mux toggle 
coverage}---that measures if the control signal to a 2:1 multiplexer expresses both 
states (0 and 1). Unlike \hwf{}, they instrument the \ac{hdl} directly, and develop 
their own custom \ac{rtl} fuzzer (Fig.~\ref{fig:hwfuzz_overview}). Unfortunately, 
\verb|RFUZZ| must be accelerated on FPGAs since coverage tracing is slow, and it is 
unclear how their \textit{mux toggle coverage} maps to existing \ac{rtl} coverage 
metrics \ac{dv} engineers care about most, e.g., code coverage and functional 
coverage~\cite{jou1999coverage,tasiran2001coverage}. Gent \etal{}~\cite{gent2016fast} propose an 
automatic test pattern generator based on custom coverage metrics, for which they 
also instrument the \ac{rtl} directly to trace. Unfortunately, like \verb|RFUZZ|, the 
scalability of their approach remains in question, given their coverage tracing method, 
and unlike \verb|RFUZZ|, they do not accelerate their simulations on FPGAs.

\subsubsection{\textbf{Design-Specific}}
Unlike the \textit{design-agnostic} approaches, the following work proposes \ac{cdg} 
techniques exclusively for processors. 
Zhang \etal{} propose \textit{Coppelia}~\cite{zhang2018end}, a tool that uses a 
custom symbolic execution engine (built on top of 
KLEE~\cite{cadar2008klee}) on software models of the \ac{rtl}.
Coppelia's goal is to target specific security-critical properties of processors;
\hwf{} enables combining such static methods with fuzzing (i.e., concolic 
execution~\cite{stephens2016driller}) for free, overcoming the limits of symbolic execution alone.
Two other processor-specific \ac{cdg} approaches are Squillero's 
\textit{MicroGP}~\cite{squillero2005microgp} and Bose \etal{}'s~\cite{bose2001genetic} that 
use a genetic algorithms to generate random assembly programs that maximize \ac{rtl} code 
coverage of a processor. Unlike \hwf{}, these approaches require custom \ac{dut}-specific
grammars to build assembly programs from.

\section{Conclusion}\label{section:conclusion}
\hwf{} is an effective solution to \ac{cdg} for hardware \ac{dv}. Unlike prior work, 
we take advantage of feature rich software testing methodologies and tools, to solve 
a longstanding problem in hardware \ac{dv}. To make our approach attractive to \ac{dv} 
practitioners, we solve several key deployability challenges, including developing 
generic interfaces (grammar \& testbench) to fuzz \ac{rtl} in a design-
agnostic manner. Using our generic grammar and testbench, we demonstrate how \hwf{} 
can achieve over 88\% \ac{hdl} code coverage of three out of four commercial-grade 
hardware designs in only one hour, with no knowledge of the \ac{dut} design or 
implementation. Moreover, compared to standard dynamic verification practices, we 
achieve over two order-of-magnitude faster design coverage with \hwf{}. 

\section*{Acknowledgment}
We thank Scott Johnson, Srikrishna Iyer, Rupert Swarbrick, Pirmin Vogel, Philipp Wagner, and other members of the OpenTitan team for their technical expertise that enabled us to demonstrate our approach on the OpenTitan IP ecosystem.

\bibliographystyle{IEEEtran}
\bibliography{sections/bibliography.bib}


\end{document}